\documentclass[acmtog]{acmart}

\usepackage{color}
\usepackage{subcaption}
\usepackage{wrapfig,graphicx}
\usepackage{natbib}

\usepackage{multirow}
\usepackage{makecell}

\begin{document}
\title{Neural Slicer for Multi-Axis 3D Printing}

\author{Tao Liu}\orcid{0000-0003-1016-4191}
\authornotemark[1]
\affiliation{
  \institution{The University of Manchester}
  \city{Manchester}
  \country{United Kingdom}
}

\author{Tianyu Zhang}\orcid{0000-0003-0372-0049}
\authornote{Joint first authors.}
\affiliation{
  \institution{The University of Manchester}
  \city{Manchester}
  \country{United Kingdom}
}

\author{Yongxue Chen}\orcid{0000-0001-6236-4158}
\affiliation{
  \institution{The University of Manchester}
  \city{Manchester}
  \country{United Kingdom}
}

\author{Yuming Huang}\orcid{0000-0001-5900-2164}
\affiliation{
  \institution{The University of Manchester}
  \city{Manchester}
  \country{United Kingdom}
}

\author{Charlie C. L. Wang}\orcid{0000-0003-4406-8480}
\authornote {Corresponding author: changling.wang@manchester.ac.uk (Charlie C. L. Wang).  }
\affiliation{
  \institution{The University of Manchester}
  \city{Manchester}
  \country{United Kingdom}
}

\authorsaddresses{Authors' addresses: Engineering Building A, The University of Manchester, Manchester, M13, United Kingdom.}

\begin{abstract}
We introduce a novel neural network-based computational pipeline as a representation-agnostic slicer for multi-axis 3D printing. This advanced slicer can work on models with diverse representations and intricate topology. The approach involves employing neural networks to establish a deformation mapping, defining a scalar field in the space surrounding an input model. Isosurfaces are subsequently extracted from this field to generate curved layers for 3D printing. Creating a differentiable pipeline enables us to optimize the mapping through loss functions directly defined on the field gradients as the local printing directions. New loss functions have been introduced to meet the manufacturing objectives of support-free and strength reinforcement. Our new computation pipeline relies less on the initial values of the field and can generate slicing results with significantly improved performance.
\end{abstract}

\ccsdesc[500]{Computing methodologies~Shape modeling}
\ccsdesc[500]{Computing methodologies~Machine learning approaches}

\keywords{curved slicing, neural network, field optimization, multi-axis motion, 3D printing}

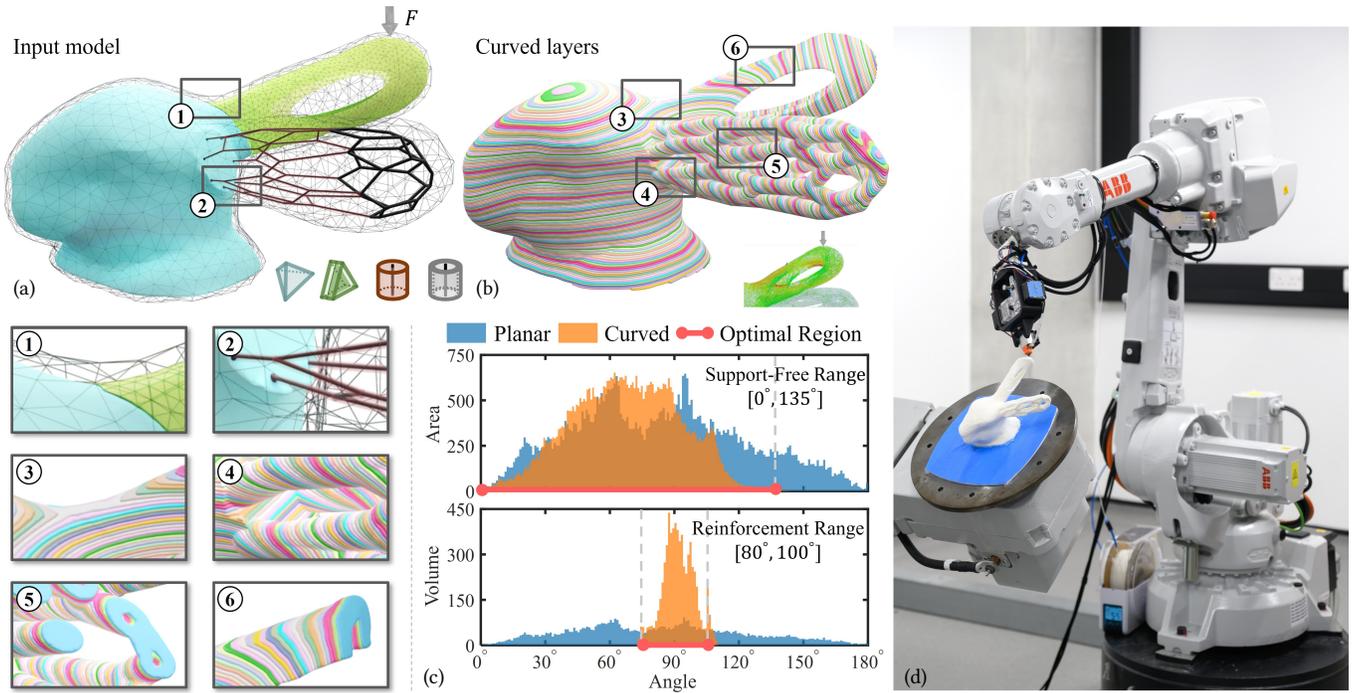
\begin{teaserfigure}
    \centering
    \includegraphics[width=\textwidth]{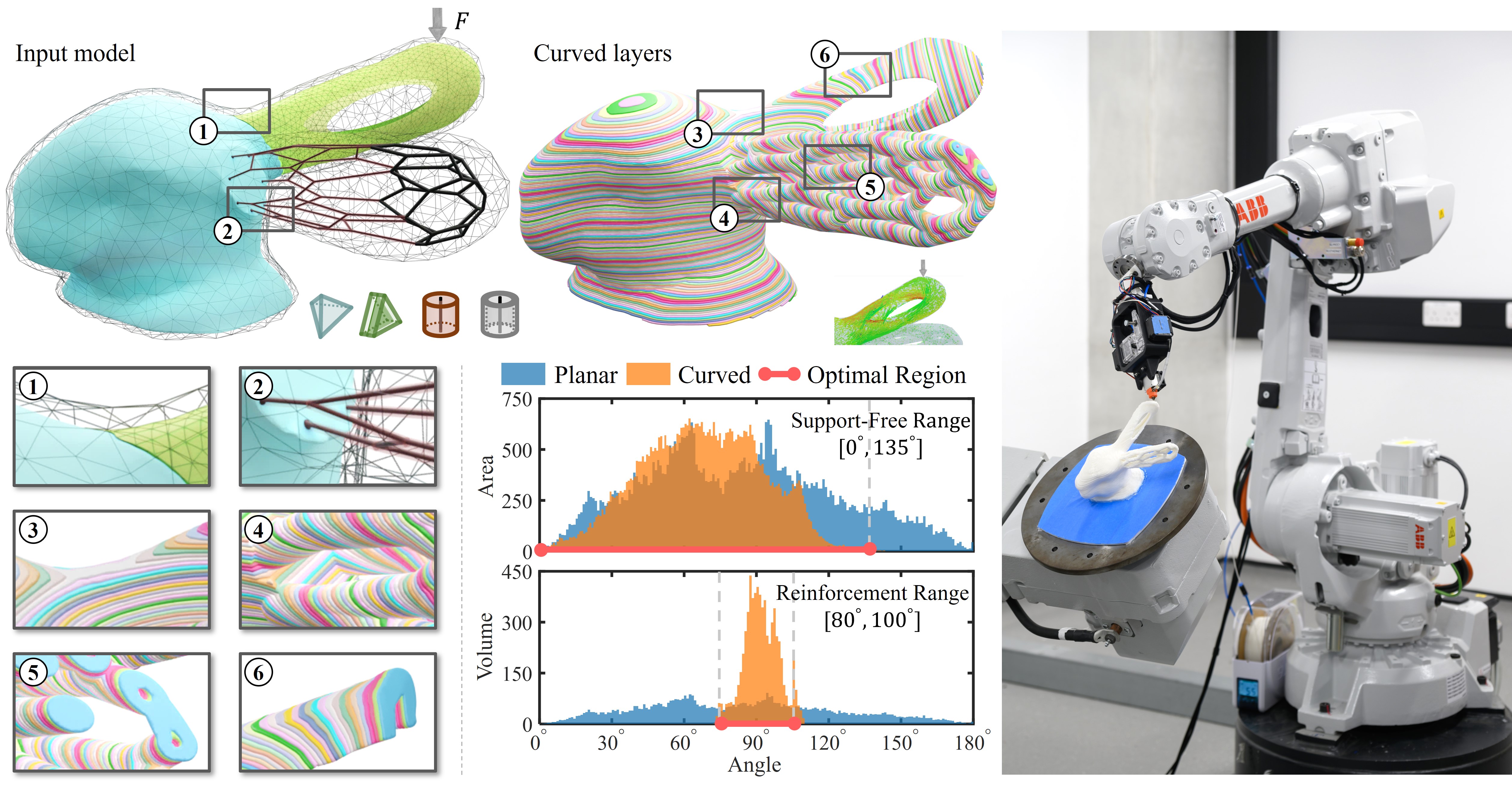}
    \put(-505,155){\small \color{black}(a)}
    \put(-330,155){\small \color{black}(b)}
    \put(-350,8){\small \color{black}(c)}
    \put(-168,8){\small \color{black}(d)}
\vspace{-5pt}
\caption{Our Neural Slicer for multi-axis 3D printing can compute curved layers for a Bunny Head model (a), incorporating diverse representations within the same model. These representations encompass a tetrahedral mesh (depicted in blue) as solid, an open surface (depicted in green) as the kernel of a shell, the skeletons (depicted in thin black lines) of struts as cylindrical solids, and also the skeletons (depicted in bold black lines) of tubular solids. (b) The curved layers are generated through the optimization of a neural network-based mapping according to the objectives of \textit{support-free} (SF) and \textit{strength reinforcement} (SR), which are formulated as the loss functions in terms of \textit{local printing directions} (LPDs). (c) The red bands in the histograms depict the required angle ranges between LPDs and other reference vectors. (d) Curved layers satisfying these requirements can be effectively generated to supervise the physical fabrication taken on a robotic system with eight \textit{degree-of-freedoms} (DOFs). The mapping is numerically computed with the assistance of a caging mesh that is independent of the discrete representations of the input model. In this example, the genus numbers of the input model and the caging mesh are $g=22$ and $g=0$ respectively.
}\label{fig:Teaser}
\end{teaserfigure}

\maketitle

\section{Introduction}
The additional degree-of-freedoms (DOFs) provided by multi-axis 3D printing offer many advantages over planar layer based printing, including the relieved need for support structures \cite{dai2018support-free}, the improved surface smoothness \cite{etienne2019curvislicer}, and the enhanced mechanical strength \cite{fang2020reinforced}. For an input model $\mathcal{M}$, these methods always define a scalar field $G(\mathbf{x})$ inside the input model $\mathcal{M}$ and extract the isosurfaces of $G(\mathbf{x})$ as curved layers for multi-axis 3D printing. Different strategies have been applied to compute an optimized field $G(\mathbf{x})$. 

\begin{figure}[t]
\centering
    \includegraphics[width=\linewidth]{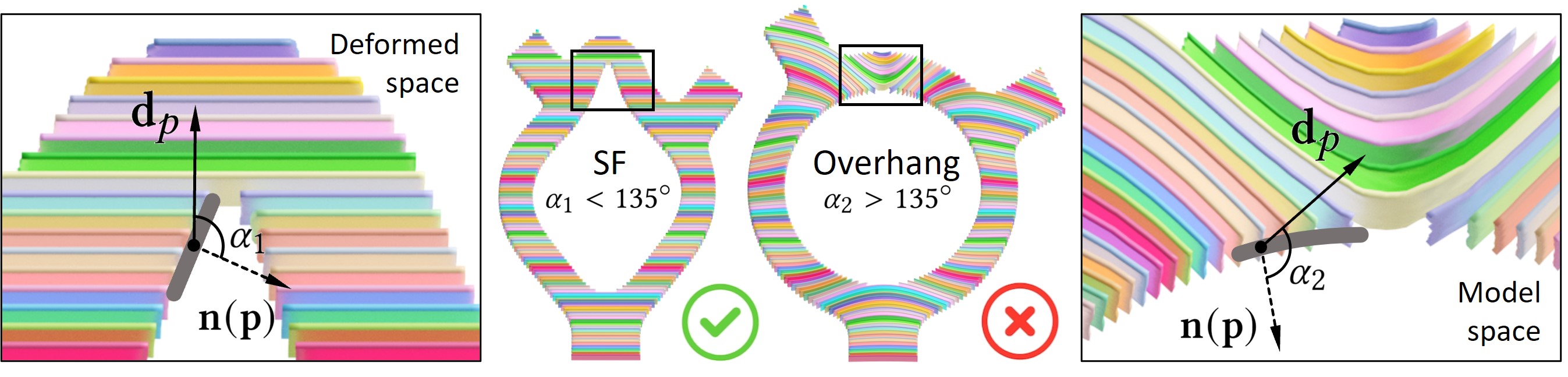}
    \put(-165,3){\small \color{black}(a)}
    \put(-124,3){\small \color{black}(b)}
\vspace{-5pt}
\caption{An illustration of the distortion caused by indirect optimization in $S^3$-Slicer: (a) the layers generated in the deformed space, and (b) curved layers in the model space obtained by the mapping. In the deformed space, the support-free requirement has been fulfilled -- i.e., the angle between the printing direction and surface normal is less than $135^\circ$. However, the angle expands when mapped back to the model space, resulting in a larger overhang area (see also the 3D printing result shown in Fig.\ref{fig:result-ring}).
}\label{fig:Problem2_Def}
\end{figure}

The most recent work of $S^3$-slicer \cite{zhang2022s3} can achieve multiple objectives on the same model by deforming the tetrahedral mesh of $\mathcal{M}$ via nonlinear optimization driven by rotations. Height values on the deformed mesh are mapped back to the input model as $G(\mathbf{x})$ and therefore obtain the curved layers. Although it can generate curved layers satisfying multiple manufacturing objectives, $S^3$-Slicer also has the following problems:
\begin{itemize}
\item The computation requires a tetrahedral mesh with high quality -- i.e., the mesh can become dense and challenging to obtain when applied to models with complex geometry and topology;

\item Objective functions of optimization are indirectly defined as rotations of elements rather than directly on the curved layers -- as a result, there is a risk of violating manufacturing requirements due to the distortion between the deformed space and the model space (see Fig.\ref{fig:Problem2_Def} for an example);

\item The success of nonlinear optimization is significantly dependent on the initial pose of an input model, which may give sub-optimal results if the initial orientation is not well-selected (see Fig.\ref{fig:result-bridge} for an example).
\end{itemize}
We propose a computational pipeline based on \textit{neural networks} (NN) to address the challenges associated with curved slicers in multi-axis 3D printing. For instance, as shown in Fig.\ref{fig:Teaser}, our slicer generates curved layers for a Bunny Head model with diverse representations and complex topology. The curved layers are optimized in accordance with manufacturing objectives of \textit{support-free} (SF) and \textit{strength reinforcement} (SR), formulated as the requirements for \textit{local printing directions} (LPDs). The resultant curved layers for the Bunny Head model have been verified through physical fabrication conducted on a robotic system with 8-DOFs.

\subsection{Our Method}
For all discrete representations, an input model $\mathcal{M}$ can always be evaluated as an implicit function $H(\mathbf{x})$, where the value defines whether a query point $\mathbf{x}$ is inside (i.e., $H(\mathbf{x})<0$) or outside the solid of $\mathcal{M}$ (i.e., $H(\mathbf{x})>0$). The smaller the absolute value $|H(\mathbf{x})|$ returned, the closer the query point $\mathbf{x}$ is to the boundary surface of $\mathcal{M}$, approximated as the zero level-set of $H(\mathbf{x})$. The function $H(\mathbf{x})$ only needs to exhibit this monotonic property near the model's boundary, rather than strictly adhering to being a distance field. We compute a continuous mapping $\lambda: \mathbf{x} \mapsto  \mathbf{y}, \forall \mathbf{x}, \mathbf{y}\in \mathbb{R}^3$ that defines the scalar field $G(\mathbf{x})$ for layer generation by the $z$-component of $\mathbf{y}$. The optimization of this mapping occurs within the model $\mathcal{M}$ (i.e., $\forall \mathbf{x} \in \mathbb{R}^3, H(\mathbf{x}) \leq 0$), and it is tailored to meet various manufacturing demands specific to multi-axis 3D printing.

We parameterize the mapping with two continuous functions, $\mathbf{q}(\mathbf{x})$ and $\mathbf{s}(\mathbf{x})$, represented as $\lambda(\mathbf{q}(\mathbf{x}),\mathbf{s}(\mathbf{x}))$, where $\mathbf{q}(\mathbf{x})$ and $\mathbf{s}(\mathbf{x})$ specify the local deformation's quaternion and scaling ratios for all $\mathbf{x} \in \mathbb{R}^3$. The functions $\mathbf{q}(\mathbf{x})$ and $\mathbf{s}(\mathbf{x})$ are modeled by neural networks with coefficients $\theta$, which are the weights associated with activation functions. The mapping $\lambda(\cdot)$ is computed numerically as a deformation applied to a volumetric mesh $\mathcal{C}$ caging the input model $\mathcal{M}$ with our effort to make the deformation differentiable. Then, the scalar field is defined as 

\begin{equation}
G(\mathbf{x}) := \mathsf{proj}_{z}\mathbf{y} = \mathsf{proj}_{z}\lambda_\theta (\mathbf{x}).
\end{equation}
Different fields are defined by different network coefficients $\theta$. 

With this new formulation, we develop a neural network-based computational pipeline to generate the optimized curved layers (i.e., the scalar field $G(\mathbf{x})$), utilizing $\theta$ as variables. This new pipeline effectively addresses all the previously mentioned challenges:
\begin{itemize}
\item The highly nonlinear nature of NN can help model more complicated deformation than those defined as piecewise functions in tetrahedral elements (i.e., the optimization is less blocked by the topology of meshes employed);

\item By establishing a differentiable neural pipeline for evaluating $G(\mathbf{x})$, we can directly define loss functions based on $G(\mathbf{x})$ and its gradients $\nabla G(\mathbf{x})$ (e.g., the LPD requirements for SF and SR) at any point in the computation domain;

\item With our novel neural pipeline, LPDs can be easily adjusted during the optimization of the scalar field $G(\mathbf{x})$, facilitated by the robustness of modern stochastic NN solvers to different initial guesses. 
\end{itemize}
We employ NN to represent continuous functions for $\mathbf{q}(\mathbf{x})$ and $\mathbf{s}(\mathbf{x})$ so that the mapping $\lambda(\mathbf{q}(\mathbf{x}),\mathbf{s}(\mathbf{x}))$ can be effectively and efficiently optimized by borrowing the computational power of the modern machine learning pipeline.

The volumetric mesh employed for caging an input model $\mathcal{M}$ is independent of $\mathcal{M}$'s discrete representation. It is an intermediate representation used in numerical computation to evaluate the mapping $\lambda$. This volumetric mesh is also employed for extracting the isosurfaces of $G(\mathbf{x})$. The resultant curved layers are obtained by trimming these isosurfaces using the implicit solid defined by $H(\mathbf{x})$.

\subsection{Contributions}
We present a neural network-based computational pipeline as a representation-agnostic slicer for multi-axis 3D printing, which makes the following technical contributions:
\begin{enumerate}
\item Formulation of the curved slicing problem as an optimization task for two continuous functions, $\mathbf{q}(\mathbf{x})$ and $\mathbf{s}(\mathbf{x})$, defining a mapping $\lambda(\mathbf{q}(\mathbf{x}),\mathbf{s}(\mathbf{x}))$, and consequently, the scalar field $G(\mathbf{x})$ for slicing.

\item Development of a differentiable pipeline of neural networks for optimization, with loss functions directly based on $\nabla G(\mathbf{x})$ (representing real LPDs), which reduces dependency on initial guesses.

\item Derivation of loss functions within the neural pipeline to address manufacturing objectives for multi-axis 3D printing, including support-free and strength reinforcement. 
\end{enumerate}
To the best of our knowledge, this is the first curved slicer that can handle models with diverse representations and complex topology. The manufacturing demands on LPDs are directly optimized, leading to improved slicing solutions, as validated through physical fabrication experiments. 
\section{Related work}\label{section:RelatedWork}
\subsection{Multi-axis 3D Printing}
Conventional planar layer based 3D printing has relatively simpler software implementation, which however introduces many limitations such the weak mechanical strength \cite{ahn2002anisotropic} and the poor surface quality \cite{chakraborty2008extruder}. Supporting structures are needed for large overhangs, leading to challenges such as hard to remove, surface damage and material waste \cite{zhang15perceptual}. With the help of additional DOFs in motion, multi-axis 3D printing offers opportunities to improve these aspects substantially. Recent developments enable the function of support-free material accumulation \cite{mitropoulou2020print, dai2018support-free, wu16printing, huang2016framefab}, enhance the mechanical strength of 3D printed parts \cite{tam20173d, fang2020reinforced}, and reduce the staircase effect \cite{etienne2019curvislicer}. Many of these approaches were implemented on robotic hardware (e.g., \cite{li2022vector, wu2020general, bhatt2021automated}) or those modified from a multi-axis CNC machine (e.g.,~\cite{fang2020reinforced, zhong2023vasco, zhang2021singular}), which have caught a lot of attention in the community of computational fabrication (e.g.,~\cite{duenser2020robocut,bartovn2021geometry}).

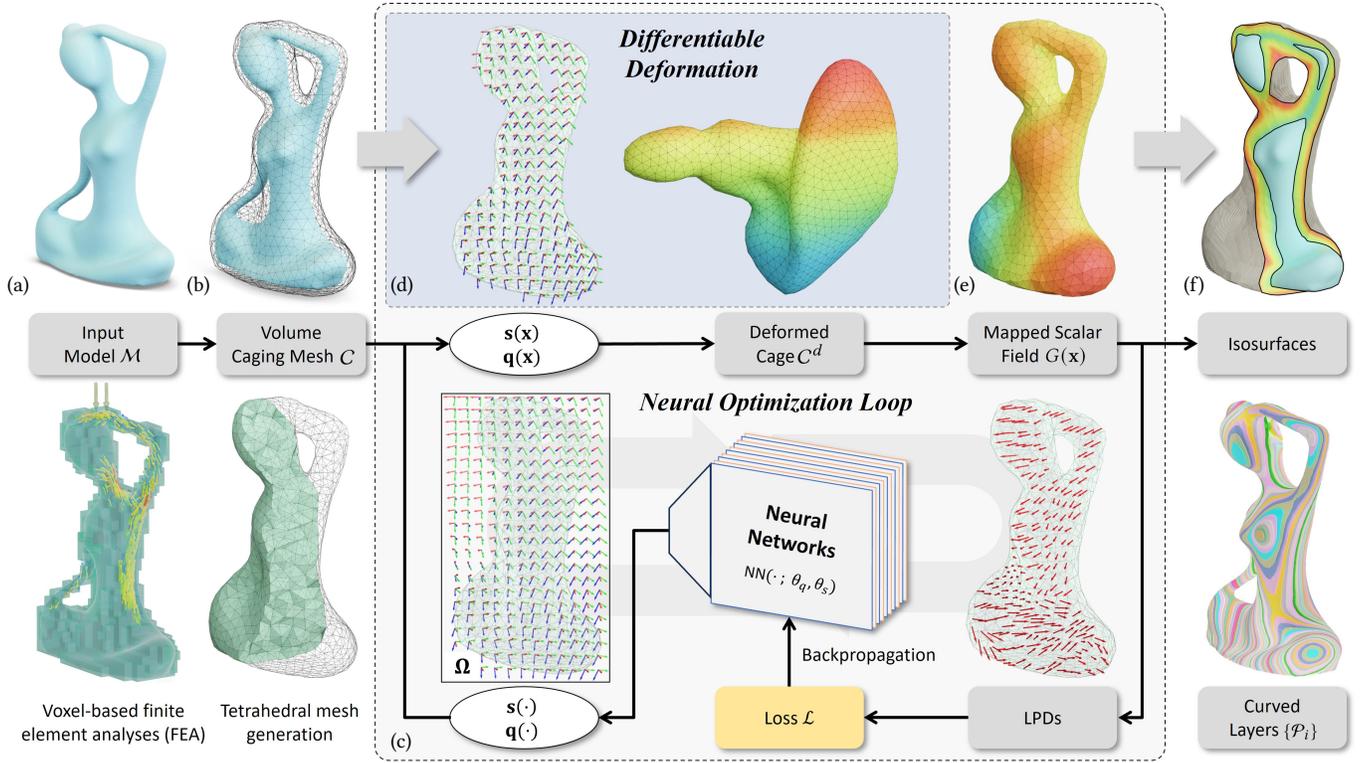
\begin{figure*}
    \centering
    \includegraphics[width=\textwidth]{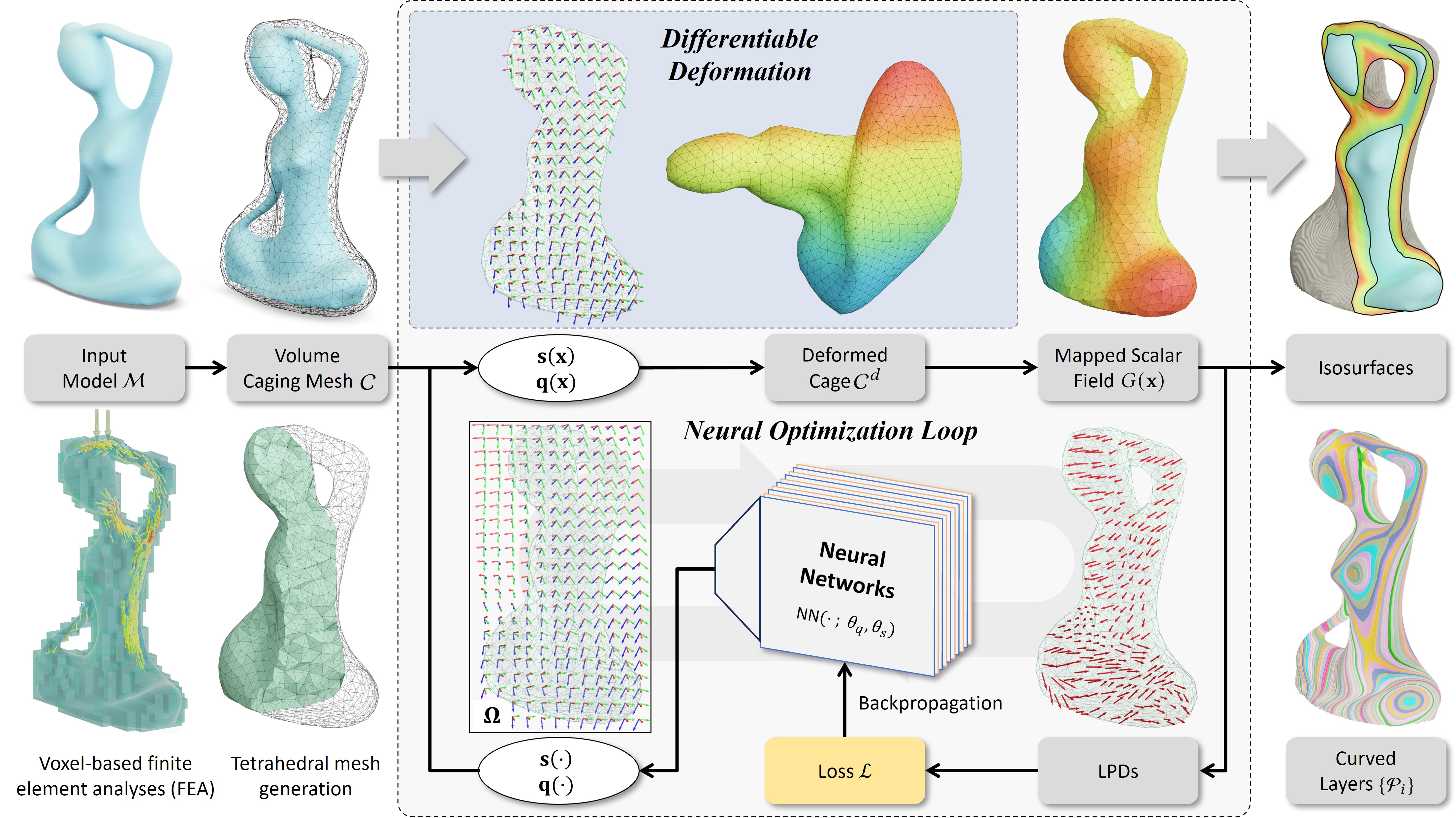}
    \put(-510,178){\small \color{black}(a)}
    \put(-442,178){\small \color{black}(b)}
    \put(-365,5){\small \color{black}(c)}
    \put(-365,178){\small \color{black}(d)}
    \put(-152,178){\small \color{black}(e)}
    \put(-65,178){\small \color{black}(f)}
\caption{An overview of our neural slicer to generate curved layers for multi-axis 3D printing. (a) The input Yoga model $\mathcal{M}$ with its implicit solid $H(\mathbf{x})$ and the distribution of principal stresses obtained from voxel-based FEA. (b) A volumetric mesh $\mathcal{C}$ caging the input model $\mathcal{M}$ is constructed, serving as the intermediate representation in numerical computation. (c) Two continuous functions $\mathbf{q}(\mathbf{x})$ and $\mathbf{s}(\mathbf{x})$ specify the quaternion and the scaling ratios of local deformation for all $\mathbf{x}\in \mathbb{R}^3$, which are represented as \textit{neural networks} (NN) to be optimized. (d) The function values of $\mathbf{q}(\mathbf{x})$ and $\mathbf{s}(\mathbf{x})$ are sampled 
to drive a differential deformation to obtain a deformed caging mesh $\mathcal{C}^d$ therefore also the mapping $\lambda(\mathbf{q}(\mathbf{x}),\mathbf{s}(\mathbf{x}))$. (e) The scalar field $G(\mathbf{x})$ is obtained from the mapping, with its gradient $\nabla G(\mathbf{x})$ serving as \textit{local printing directions} (LPDs) that play a pivotal role in formulating the loss functions utilized for optimization. (f) After computing an optimized $G(\mathbf{x})$, its isosurfaces are extracted on the caging mesh $\mathcal{C}$ and trimmed by the implicit solid $H(\mathbf{x})$ to obtain the curved layers.
}\label{fig:overview}
\end{figure*}

Scalar fields have been widely employed to define curved layers inside a solid by the isosurfaces. For instance, Dai et al.~\shortcite{dai2018support-free} sought the help of voxel representation to progressively compute a material accumulation field while keeping a convex-front as a conservative method to prevent potential collisions. To improve the surface quality of 3D printed models, Etienne et al.~\shortcite{etienne2019curvislicer} developed a height-field like deformation to generate slightly curved layers for 3-axis machines. Fang et al.~\shortcite{fang2020reinforced} presented a pipeline exploiting the anisotropy of mechanical properties in fused filament fabrication with the help of vector fields. They first optimized a vector field to align with the minimal stresses, and later computed a scalar field with its gradients approximating the vector field. The gradients of scalar fields are not directly optimized. Similar indirect problem exists in the recent work of $S^3$-Slicer \cite{zhang2022s3}. Moreover, many of these methods require a volumetric mesh with high quality, which is challenging for models with complex geometry. 

\subsection{Field-based Optimization}
The challenge of achieving different objectives in multi-axis 3D printing can be formulated as a problem of optimizing LPDs at different points, essentially forming vector fields. A possible approach to model LPDs is to extend the techniques for processing vector fields defined on mesh surfaces (e.g., \cite{de2016vector}) to volumetric meshes. Another option is frame-based field optimization \cite{huang2011boundary, li2012all, ray2016practical}, which has been used to tackle other problems of computational fabrication (e.g., \cite{montes2023differentiable, mitra2023helix}). However, these methods can be very time-consuming, as evidenced by the work presented in \cite{arora2019volumetric} for volumetric Michell trusses.

Recently, neural network-based \textit{implicit neural representation} (INR) has become very popular because of its capability to effectively solve problems in reconstruction and optimization \cite{de2023deep}. INRs are usually formulated as neural networks in a differentiable pipeline of computation and determined by self-learning based optimization using a NN solver (e.g., Adams~\cite{kingma2014adam}). Example works include DeepSDF \cite{park2019deepsdf} that approximates continuous signed distance functions for surface reconstruction and SIRENs \cite{sitzmann2020implicit} that introduces periodic activation functions for modeling complex natural signals and solving partial differential equations. 

With the help of INR, recent effort has been made to overcome the limitations caused by the discrete nature of meshes in deformation. Methods such as \cite{groueix20183d, groueix2019unsupervised, huang2020meshode, jiang2020shapeflow, yang2021geometry} utilized the implicit vector fields to represent the deformation, which can be applied to models in different representations. The optimized deformation of a template model can be computed by optimizing INR according to different objectives and constraints. This strategy has been adopted to solve problems of non-rigid registration \cite{deng2021deformed, sundararaman2022reduced}, surface reconstruction  \cite{williams2019deep} and parameterization \cite{morreale2021neural,aigerman2022neural}. However, the translation based nature of implicit vector fields does not provide rotational-invariant deformation. In this paper, we employ NNs as continuous functions for quaternions and scaling ratios to determine optimized curved layers for multi-axis 3D printing. 

\subsection{Mesh-based Deformation and Parameterization}
Deforming a given model is usually formulated as an optimization problem (e.g., \cite{sorkine2009interactive}). Earlier approaches mainly defined the geometric structures according to the local connectivity, where a surface mesh or a volumetric mesh with high quality is needed. To overcome this challenge, cage-based deformation (e.g., \cite{yifan2020neural,hu2015support}) has been employed to handle wild geometry such as scattered points or polygon soups. 

We proposed to construct a `cage'-like tetrahedral mesh, which encloses the input model but has completely different connectivity, as the intermediate discretization for computing the mapping. The scale-controlled deformation proposed in \cite{zhang2022s3} is employed in our work to compute the mapping. Specifically, the as-rigid-as-possible (ARAP) energy \cite{sorkine2007rigid, sorkine2004laplacian} is defined as the function to be minimized while allowing to scale with specified ratios along three orthogonal directions. The ARAP energy can be efficiently minimized by a local/global scheme \cite{bouaziz2012shape,liu2008local}.
Although the effort of differentiation has been made in \cite{bacher2014spinit,romain2013make}, ARAP deformation has not been employed in the NN-based computational pipeline -- especially with respect to the fields of quaternions and scaling ratios.

Our work to compute an optimized mapping shares certain similarities with the volumetric parameterization problems \cite{patane2013surface}. Specifically, the task of computing a mapping for the SF and SR demands can be viewed as a specialized volumetric parameterization, incorporating orientation constraints. Different from the large literature of parameterization that mainly focus on resolving foldovers to ensure local injectivity (ref.~\cite{du2020lifting, fu2015computing, garanzha2021foldover, kovalsky2015large, liao2021real, schuller2013locally, su2019practical, rabinovich2017scalable} ), local injectivity is not particularly required in our problem. 

\section{Neural Optimization for Slicing}\label{SecNeuralSlicerPipeline}
We introduce the computational pipeline of our Neural Slicer in this section. First of all, a neural network-based formulation is introduced for the mapping $\lambda(\cdot)$ to be optimized. After that, the steps of our slicing algorithm are presented. An overview illustration is given in Fig.\ref{fig:overview}. 

\subsection{Differentiable Mapping as Neural Network}\label{subsecDiffMapFunc}
The basic idea of our approach is to construct a differentiable mapping  $\lambda(\cdot) \in \mathbb{R}^3$ by neural networks. The scalar field for layer generation can be defined by the $z$-component of $\lambda(\cdot)$ as $G(\mathbf{x})$, where the function $\lambda(\cdot)$ is parameterized with the network coefficients $\theta$. Therefore, the scalar field $G(\mathbf{x})$ for curved layer generation can be optimized by self-learning according to the manufacturing requirements directly defined on $G(\mathbf{x})$ and $\nabla G(\mathbf{x})$ as loss functions.

Simply defining the mapping $\lambda(\cdot)$ by a neural network without any geometric meaning leads to a learning process that converges very slowly. We propose to formulate $\lambda(\cdot)$ as $\lambda(\mathbf{q}(\mathbf{x}),\mathbf{s}(\mathbf{x}))$ by two continuous functions $\mathbf{q}(\mathbf{x}) \in \mathbb{R}^4$ and $\mathbf{s}(\mathbf{x}) \in \mathbb{R}^3$ that define the local rotation (as quaternion) and the local scaling ratios at any point $\mathbf{x} \in \mathbb{R}^3$. $\mathbf{q}(\mathbf{x})$ and $s(\mathbf{x}) \in \mathbb{R}^3$ are represented by two neural networks with coefficients $\theta_q$ and $\theta_s$ respectively. $\lambda(\cdot)$ are then computed from $\mathbf{q}(\mathbf{x})$ and $\mathbf{s}(\mathbf{x})$ in a discrete manner. The space $\Omega$ around the input model $\mathcal{M}$ is discretized into a volumetric mesh $\mathcal{C}$ with tetrahedral elements, where $\mathcal{C}$ encloses the solid of $\mathcal{M}$ as a cage. The scale-controlled ARAP deformation \cite{zhang2022s3} is employed\footnote{Note that the deformation is computed in a space $\Omega$ instead of the input model to determine the mapping.} to compute a deformed caging mesh $\mathcal{C}^d$ so that the quaternions and the scaling ratios can be approximated at every element of $\mathcal{C}$. Here we use the centers of elements as the sample points. With the help of barycentric coordinates, every point $\mathbf{x} \in \Omega$ is mapped to a new position $\mathbf{y} \in \Omega^d$ in the new space $\Omega^d$ defined around $\mathcal{C}^d$. A discrete form of the mapping  $\lambda(\cdot)$ is obtained, the $z$-component of which is the scalar field  $G(\mathbf{x})$. By deriving the formulas to compute $\partial G / \partial \theta_q$ and $\partial G / \partial \theta_s$ w.r.t. the positions of vertices on $\mathcal{C}^d$ (details can be found in Sec.~\ref{subsecDiffDeform}), a mapping $\lambda(\cdot)$ can be optimized via the differentiable pipeline of neural networks. 

\subsection{Slicing Algorithm}\label{subsecAlgorithmOverview}
The algorithm consists of three major stages: pre-processing (Steps 1-5), mapping optimization (Steps 6-10), and post-processing (Step 11). The steps are presented as follows (see the illustration in Fig.\ref{fig:overview}):
\begin{enumerate}
\item Construct the solid $\mathcal{H}$ as an implicit function $H(\mathbf{x})$ for the input model $\mathcal{M}$;

\item Apply voxel-based FEA to compute stress field inside the solid $H(\mathbf{x}) \leq 0$;

\item Generate a tetrahedral mesh $\mathcal{C}$ as the intermediate discretization of the space $\Omega$ enclosing $\mathcal{H}$ (Sec.~\ref{subsecCageGeneration});

\item Build a set of sample points as $\mathcal{B}$ on the surface boundary of $\mathcal{M}$, which is employed to evaluate the loss functions for SF requirements;

\item Build a set of sample points as $\mathcal{T}$ at the center of voxels for evaluating the strength reinforcement loss in the relevant region -- i.e., the region of the top 10\% maximal stresses; 

\item Initialize the neural networks for $\mathbf{q}(\mathbf{x})$ and $\mathbf{s}(\mathbf{x})$;

\item Compute a deformed mesh of $\mathcal{C}$ as $\mathcal{C}^d$ by the scale-controlled ARAP deformation;

\item Evaluate the differentiation of the deformation to prepare for backpropagation (Sec.~\ref{subsecDiffDeform});

\item Update the networks of $\mathbf{q}(\mathbf{x})$ and $\mathbf{s}(\mathbf{x})$ by the NN solver according to the loss functions (Sec.~\ref{secLossFunctions});

\item Go back to step (7) until the learning converges;

\item Extract the isosurfaces of $G(\mathbf{x})$ from the volumetric mesh $\mathcal{C}$ and trim them by the implicit solid $H(\mathbf{x}) \leq 0$ (Sec.~\ref{subsecSlicingOnCage}).
\end{enumerate}
As the mapping is computed on an intermediate mesh $\mathcal{C}$, our slicing algorithm does not require a perfect volumetric mesh as the discretization of the input model $\mathcal{M}$. 
\section{Loss Functions}\label{secLossFunctions}
In order to optimize a mapping function that leads to curved layers satisfying different manufacturing requirements, loss functions are defined in this section for strength reinforcement, support-free, and collision avoidance. All loss functions are defined according to the LPDs -- i.e., $\mathbf{d}_p = {\nabla G(\mathbf{x})} / {\| \nabla G(\mathbf{x}) \|}$ w.r.t. the scalar field $G(\mathbf{x})$ to be optimized. Harmonic losses of $\mathbf{q}(\mathbf{x})$ and $\mathbf{s}(\mathbf{x})$ are introduced for the purpose of regularization. 

\subsection{Strength Reinforcement}\label{subsecSRLoss}
\begin{wrapfigure}[6]{r}{0.22\linewidth}
\begin{center}
\hspace{-18pt}\includegraphics[width=1.0\linewidth]{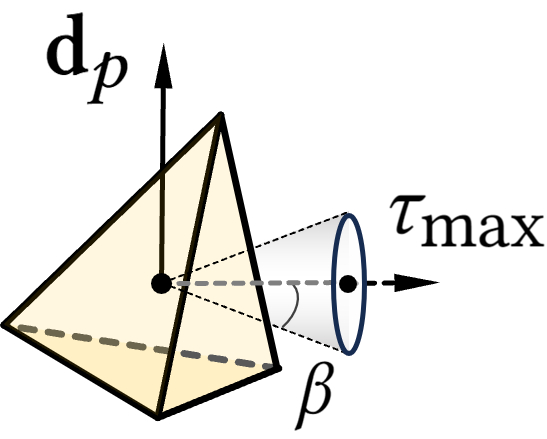}
\end{center}
\end{wrapfigure} 
Anisotropic mechanical behavior is a prominent characteristic observed in models fabricated through filament-based 3D printing, primarily due to the limited adhesion between deposited filaments \cite{riddick2016fractographic}. Under specific load conditions, the mechanical strength of printed models can be significantly enhanced when the filaments are aligned along the distribution of principal stresses \cite{tam2017additive}. Specifically, the 3D printed model can have a reinforced strength when the LPDs are nearly perpendicular to directions of the maximal principal stresses $\mathbf{\tau}_{\max}$ as $|\mathbf{d}_p \cdot \mathbf{\tau}_{\max}| \leq \sin \beta$ (ref. \cite{fang2020reinforced}). According to this criterion, we define the following loss function evaluated at the sample points $\mathbf{p} \in \mathcal{T}$ (i.e., the centers of voxels in the relevant region) for the purpose of strength reinforcement.
\begin{equation}\label{eqSRLoss}
\mathcal{L}_{SR}:=\sum_{\mathbf{p}\in \mathcal{T}} |V_{e}| ~\sigma \left( k_{SR} (|\mathbf{d}_p \cdot \mathbf{\tau}_{\max}(\mathbf{p}) | -\sin\beta ) \right)
\end{equation}
where $\sigma(\cdot)$ is the sigmoid function and $|V_{e}|$ is the volume of each voxel element for FEA. $\beta$ is a parameter that introduces a level of tolerance and is related to material properties and printing temperature. $k_{SR}$ is the logistic growth rate to tune the `steepness' of the loss function. By experimental tests, $\beta=10^{\circ}$ and $ k_{SR}=15$ are employed for all examples in this paper.

\subsection{Support-Free}\label{subsecSFLoss}
Two different criteria are considered to define the loss functions to optimize the curved layers to enable \textit{support-free} (SF) 3D printing based on 1) surface normal and 2) point extremity respectively. All are evaluated on the sample points $\mathbf{p} \in \mathcal{B}$. Note that $\mathcal{B}$ can exclude points in the regions that have been naturally supported by the printing platform (e.g., those on the bottom plate of the Bunny Head model). 

\begin{wrapfigure}[8]{r}{0.35\linewidth}\vspace{-15pt}
\begin{center}
\hspace{-18pt}\includegraphics[width=1.0\linewidth]{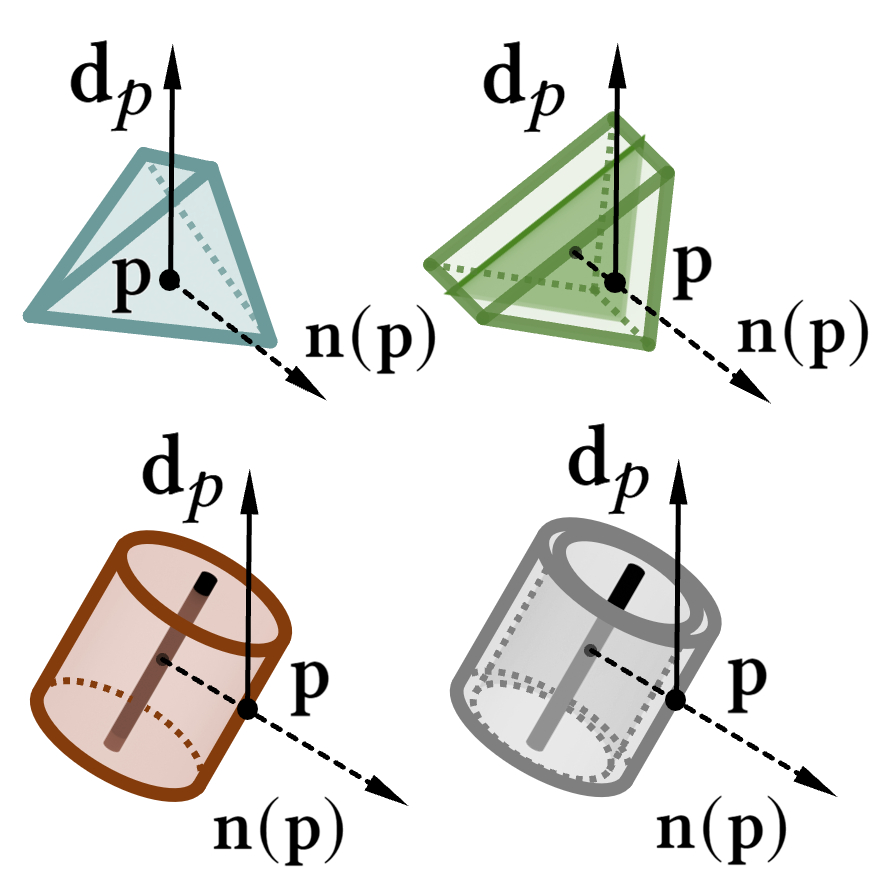}
\end{center}
\end{wrapfigure} 
Given an implicit function $H(\mathbf{x})$, the surface normal at a sample point $\mathbf{p} \in \mathcal{B}$ can generally be evaluated by the implicit function as $\mathbf{n}(\mathbf{p}) = - \nabla H(\mathbf{p}) / \| \nabla H(\mathbf{p}) \|$. Specifically, $\mathbf{n}(\mathbf{p})$ can be precisely obtained from the input model or the skeleton of the convolution solid (including the shell, the cylindrical solid, and the tubular solid) as shown above. In the context of multi-axis 3D printing, the SF requirement for LPDs and surface normals is defined according to the local self-supporting angle $\alpha$ as $- \mathbf{n}(\mathbf{p}) \cdot \mathbf{d}_p  \leq \sin \alpha$ (ref.~\cite{zhang2022s3}). Therefore, we define the following loss function.
\begin{equation}\label{eqSFLoss}
\mathcal{L}_{SF}:=\sum_{\mathbf{p}\in \mathcal{B}} |A_\mathbf{p}| ~\sigma \left( k_{SF} (- \mathbf{n}(\mathbf{p}) \cdot \mathbf{d}_p -\sin\alpha ) \right)
\end{equation}
where $\sigma(\cdot)$ is the sigmoid function and $k_{SF} = 30$ was used as the logistic growth rate for this loss function. $|A_\mathbf{p}|$ is a weight reflecting the surface area covered by a sample point, which can be estimated by the average squared distance between $\mathbf{p}$ and its $k$-nearest neighbors. We use $k=10$ for all our examples. 

\begin{wrapfigure}[8]{r}{0.4\linewidth}
\begin{center}
\hspace{-15pt}\includegraphics[width=1.0\linewidth]{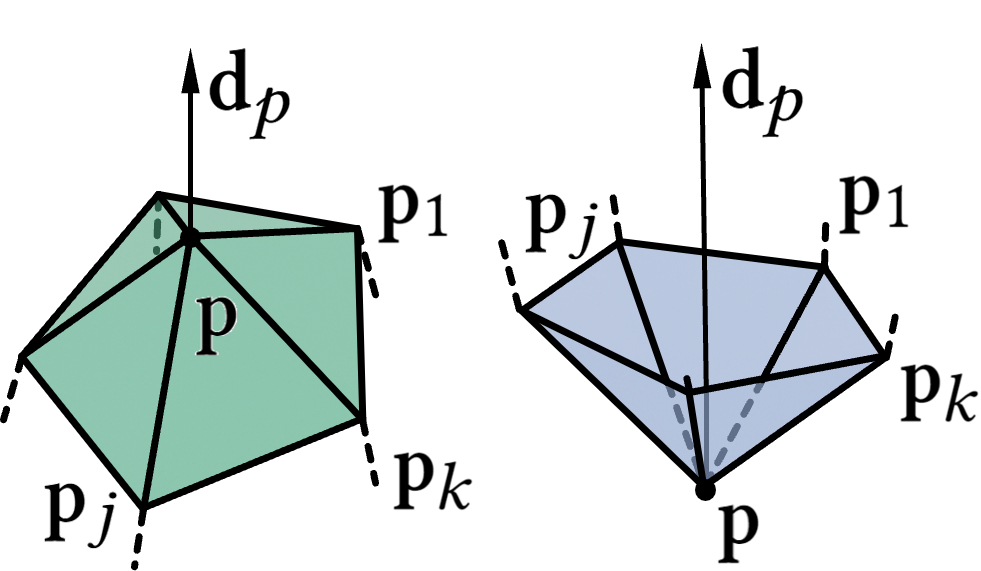}
\end{center}
\end{wrapfigure} 
Similar to the scenario of point overhang discussed in \cite{vanek2014clever}, we need to prevent making a surface point as local minimal with reference to the LPD $\mathbf{d}_p$. Specifically, giving $\mathcal{N}_{\mathbf{p}}$ as the set of $\mathbf{p}$'s $k$-nearest neighbors, $\mathbf{p}$ is a local minimum (i.e., point overhang) when 
\begin{center}
$(\mathbf{p}_j - \mathbf{p}) \cdot \mathbf{d}_p > 0$ \hspace{10pt} ($\forall \mathbf{p}_j \in \mathcal{N}_{\mathbf{p}}$).
\end{center} 
The following loss is defined to avoid generating LPDs that lead to local `minimum' on boundary surfaces (i.e., point overhang).
\begin{equation}\label{eqPOLoss}
\mathcal{L}_{PO}:=\sum_{\mathbf{p}\in \mathcal{B}} |A_\mathbf{p}| \max (0, \min_{\mathbf{p}_j \in \mathcal{N}_{\mathbf{p}}} ( (\mathbf{p}_j - \mathbf{p}) \cdot \mathbf{d}_p ))
\end{equation}
where $\min(\cdot)$ function is implemented by the min pooling layer and $\max(0, \cdot)$ can be easily realized by the ReLU function. Again, $|A_\mathbf{p}|$ is the area weight of a sample point. 

\subsection{Collision Avoidance}
\begin{wrapfigure}[8]{r}{0.5\linewidth}
\begin{center}
\hspace{-25pt}\includegraphics[width=1.0\linewidth]{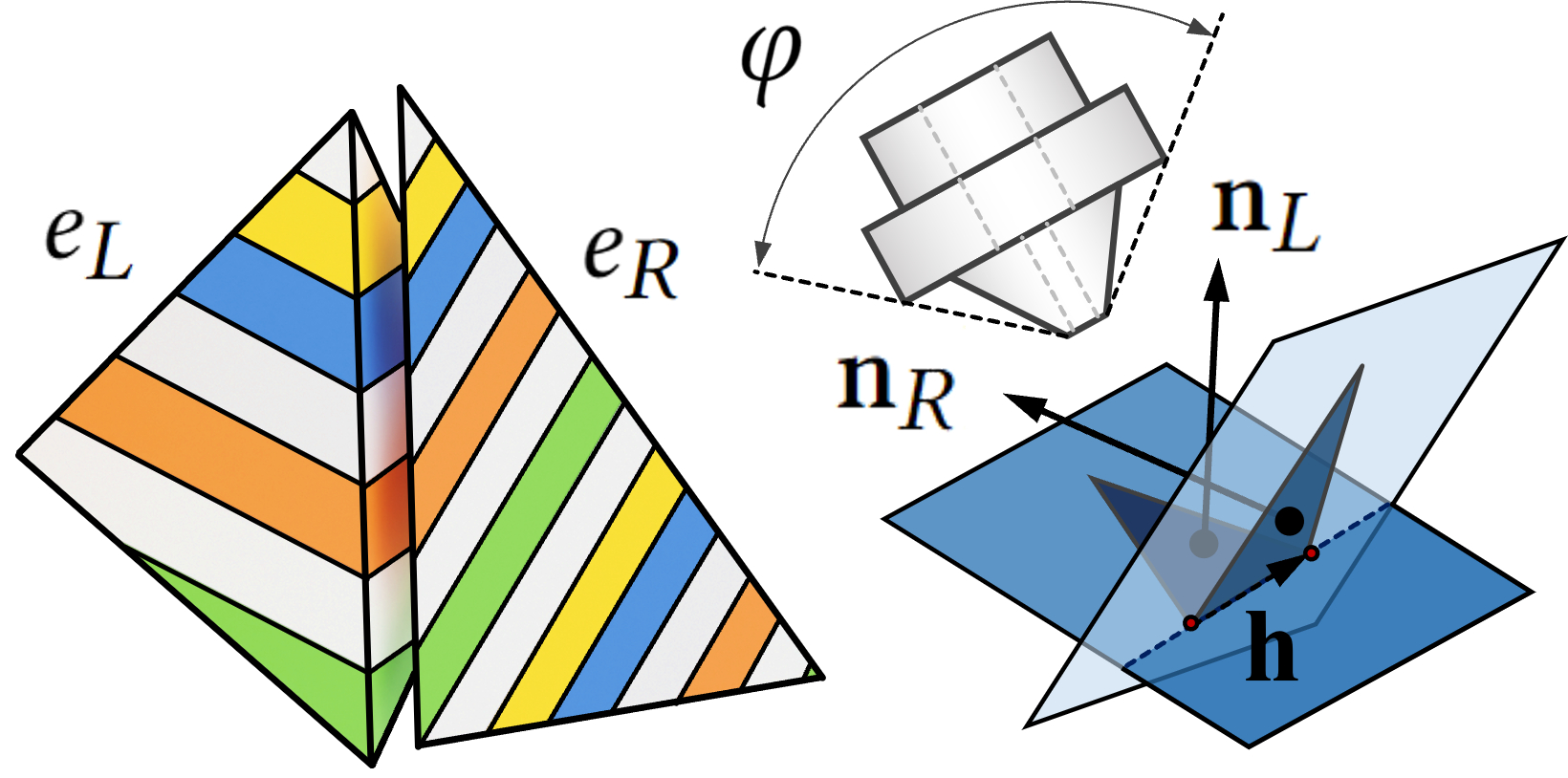}
\end{center}
\end{wrapfigure} 
As more DOFs of motion are enabled for the process of multi-axis 3D printing, the collision can happen locally between the working surface and the printer head if the curved layers are not designed properly. As shown on the right, the geometry of a printer head forms a cone shape with the apex angle $\varphi$. When using this cone to touch a point on the working surface, local collision will occur in any concave region with a dihedral angle less than $\varphi$. 

When the scalar field $G(\mathbf{x})$ is defined in a piecewise linear manner on the elements of the caging mesh $\mathcal{C}$, the normal of isosurfaces in an element $e$ is a constant vector as $\mathbf{n}_e = \nabla G(\mathbf{x}) / \| \nabla G(\mathbf{x}) \|$. The concavity of isosurfaces in two neighboring elements can be evaluated by $(\mathbf{n}_L \times \mathbf{n}_R) \cdot \mathbf{h}$, where $\mathbf{n}_L$ and $\mathbf{n}_R$ are the normal vectors in the left and the right elements $e_L$ and $e_R$ and $\textbf{h}$ is the unit vector obtained from the isoline of $G(\mathbf{x})$ defined on the triangle shared by two elements. Borrowing the study of gouging (i.e., local collision) in multi-axis CNC machining \cite{bartovn2021geometry}, the local concavity needs to satisfy the following condition to be collision-free
\begin{equation} \nonumber
    \left\{  
        \begin{array}{rl}
             -\sin{\varphi} \leq (\mathbf{n}_L \times \mathbf{n}_R) \cdot \mathbf{h} <0 & (\varphi \leq \frac{\pi}{2}),  \\  
             (\mathbf{n}_L \times \mathbf{n}_R) \cdot \mathbf{h} \leq -\sin{\varphi} & (\varphi > \frac{\pi}{2}).
        \end{array}  
    \right.  
    \label{eq:collision}
\end{equation}
This can be further formulated into a loss as
\begin{equation}\label{eqCollisionLoss}
\begin{aligned}
\mathcal{L}_{CA}:= \sum_{(e_L,e_R)\in\mathcal{N}_{e}} 
 & \max(0,(\mathbf{n}_L\times\mathbf{n}_R)\cdot\mathbf{h}) \\
 & + \max(0,-(\mathbf{n}_L\times\mathbf{n}_R)\cdot\mathbf{h}-\sin(\varphi))                            
\end{aligned}
\end{equation}
for a relative sharp printer head with $\varphi \leq \frac{\pi}{2}$. Or we define the loss function as follows
\begin{equation}\label{eqCollisionLoss2}
\mathcal{L}_{CA}:= \sum_{(e_L,e_R)\in\mathcal{N}_{e}} \max(0,(\mathbf{n}_L\times\mathbf{n}_R)\cdot\mathbf{h}+\sin(\varphi))
\end{equation}
for a relatively flat printer head with $\varphi > \frac{\pi}{2}$. Again, $\max(0, \cdot)$ is realized by the ReLU function in our implementation.
 
Different from previous loss functions, $\mathcal{L}_{CA} = 0$ needs to be processed as a hard constraint to ensure collision-free. Details can be found in Sec.~\ref{subsecNNArchOptm}. This collision avoidance loss only considers the local collision. Global collision is checked and resolved during the motion planning process of the physical realization (ref. \cite{zhang2021singular,ezair2018volumetric}), which is not the focus of this paper.

\subsection{Harmonics}\label{subsecHarmonics}
Harmonic losses are employed to control the smoothness of $\mathbf{q}(\mathbf{x})$ and $\mathbf{s}(\mathbf{x})$ to be learned. Let $\mathcal{N}_e$ define the set of neighboring element pairs in the caging mesh $\mathcal{C}$, the differences of scale ratios and quaternions are measured on pairs of neighboring elements.  

First, a loss is introduced to minimize the difference of scaling ratios $\mathbf{s}(\cdot)$ so that avoids the radical change of layer thickness in neighboring regions.
\begin{equation}\label{eqScaleHarmonicLoss}
    \mathcal{L}_{HS} := \sum_{(e_i, e_j)\in \mathcal{N}e} \frac{1}{2}(|V_{e_i}|+|V_{e_j}|) \| \mathbf{s}(\mathbf{c}_{e_i}) - \mathbf{s}(\mathbf{c}_{e_j})\|^2
\end{equation}
where $\mathbf{c}_{e_i}$ and $\mathbf{c}_{e_j}$ are the centers of the neighboring elements $e_i, e_j$, and $|V_{e_i}|$ and $|V_{e_j}|$ are the volumes of the elements. The scaling ratios are evaluated on the NN-based continuous function $\mathbf{s}(\cdot)$.

Furthermore, it is important to generate curved layers with smoothly varied normals to ensure smooth and continuous motion of the printer head during fabrication. We control this property of resultant curved layers by introducing a loss that evaluates the difference of quaternions in neighboring regions. 
\begin{equation}\label{eqQuaternionHarmonicLoss}
\mathcal{L}_{HQ} := \sum_{(e_i, e_j)\in \mathcal{N}e} \frac{1}{2}(|V_{e_i}|+|V_{e_j}|) \left( 1 - \frac{\mathbf{q}(\mathbf{c}_{e_i})}{\| \mathbf{q}(\mathbf{c}_{e_i}) \|} \cdot \frac{\mathbf{q}(\mathbf{c}_{e_j})}{\| \mathbf{q}(\mathbf{c}_{e_j}) \|} \right)^2
\end{equation}
where $\mathbf{q}(\cdot)$ is the NN-based continuous function of quaternions. 

Among all metrics for 3D rotations \cite{huynh2009metrics}, we select one with a simple form that is normalized to bound the loss value.

\subsection{Total Loss}
In summary, the curved layer generation problem can be formulated as a learning-based optimization process to minimize the total loss:
\begin{equation}
    \mathcal{L} = w_1 \mathcal{L}_{SF} + w_2 \mathcal{L}_{SR} + w_3 \mathcal{L}_{OP} + \mathcal{L}_{HS} + \mathcal{L}_{HQ}, 
\end{equation}
where $w$'s are the balancing weights for different loss terms. Considering the collision avoidance loss as a hard constraint, that is 
\begin{equation}
    \arg \min_{\theta_q, \theta_s}  \mathcal{L}   \quad s.t. \; \mathcal{L}_{CA} = 0.
\end{equation}

\section{Implement Details}
\subsection{Differentiable Deformation as Mapping}\label{subsecDiffDeform}

In our current implementation, the mapping $\lambda(\cdot)$ is computed on the volumetric caging mesh $\mathcal{C}$ by using the scale-controlled ARAP deformation \cite{zhang2022s3}. That is to determine the new position $\mathbf{v}^d$ for every vertex $\mathbf{v} \in \mathcal{C}$ so that the quaternions and the scale ratios given by the functions $\textbf{q}(\cdot)$ and $\textbf{s}(\cdot)$ are satisfied in the elements of $\mathcal{C}$ via an optimization formulation. We provide the details of this deformation below and then analyze its differentiability. 

For every element $e \in \mathcal{C}$, we determine its rotation matrix $\mathbf{R}_e$ by the quaternion $\mathbf{q}(\mathbf{c}_e)$ at the element's center $\mathbf{c}_e$. Its scale matrix $\mathbf{S}_e = \mathrm{diag}(\mathbf{s}^x(\mathbf{c}_e),\mathbf{s}^y(\mathbf{c}_e),\mathbf{s}^z(\mathbf{c}_e))$ is also computed by $\textbf{s}(\cdot)$ at the center $\mathbf{c}_e$. Then, the locally scaled and rotated element $e$ can be computed by $\mathbf{R}_e\mathbf{S}_e(\mathbf{NV}_e)^\mathrm{T}$ with $\mathbf{V}_e$ being a position matrix formed by the coordinates of $e$'s four vertices. $\mathbf{N}$ is used to transfer an element’s center to the origin as $\mathbf{N}=\mathbf{I}_{4\times4}-\frac{1}{4}\mathbf{1}_{4\times4}$. We compute the deformed caging mesh $\mathcal{C}^d$ as
\begin{equation}\label{eqARAPDeform}
\arg \min_{\mathcal{C}^d} \underbrace{\sum_{e\in\mathcal{C}} \|(\mathbf{NV}_e^d)^\mathrm{T} -  \mathbf{R}_e \mathbf{S}_e (\mathbf{NV}_e)^\mathrm{T} \|_F^2}_{\mathrm{Position-Compatibility}} + \underbrace{\gamma \sum_{\mathbf{v} \in\mathcal{C}} \|\mathbf{v}^d - \mathbf{v} \|^2}_{\mathrm{Regularization}}
\end{equation}
where $\| \cdot \|_F$ is the Frobenius norm, and the positions of vertices in $\mathcal{C}^d$ to be determined are kept in the position matrix $\mathbf{V}_e^d$. The regularization term is added to improve the conditional number of the linear system for solving this problem. 

Denoting $\mathbf{\xi}^0=\{ \mathbf{v} \}$ and $\mathbf{\xi} = \{ \mathbf{v}^d \}$ as the vectors containing the positions of all vectors before and after deformation, the above equation is actually in the regularized least-squares form as
\begin{equation}\label{eqARAPDeform2}
\arg \min_{\mathbf{\xi}} \| \mathbf{A} \mathbf{\xi} - \mathbf{b} \|^2 + \gamma \| \mathbf{\xi} - \mathbf{\xi}^0\|^2.
\end{equation}
with $\mathbf{A}$ and $\mathbf{b}$ derived from the the position compatibility term of Eq.(\ref{eqARAPDeform}). Then, the positions of the deformed mesh are obtained as 
\begin{equation}\label{eqARAPDeformSolution}
\mathbf{\xi} = (\mathbf{A}^{T}\mathbf{A}+\gamma \mathbf{I})^{-1}(\mathbf{A}^{T}\mathbf{b}+ \gamma  \mathbf{\xi}^0)
\end{equation}
with $\mathbf{I}$ as an identity matrix.

Given a query point $\mathbf{x} \in \mathbb{R}^3$ contained by an element of $\mathcal{C}$ with four vertices $\mathbf{v}_i$ ($i=1,\ldots,4$), we can determine $\mathbf{x}$'s barycentric coordinate as $a_i(\mathbf{x})$ with $\sum_i a_i(\mathbf{x}) \equiv 1$ and $a_i(\mathbf{x}) \in [0,1]$. As a result, the mapping $\lambda(\cdot)$ is obtained as $\mathbf{y} = \sum_i a_i(\mathbf{x}) \mathbf{v}^d_i$. The scalar field $G(\mathbf{x})$ as the $z$-components of $\mathbf{y}$ is 
\begin{equation}\label{eqMappingByBarycentricCoord}
    G(\mathbf{x}) = \mathrm{diag}(0,0,1) \sum_i a_i(\mathbf{x}) \mathbf{v}^d_i.
\end{equation}

This scalar field $G(\mathbf{x})$ determined by the NN-based mapping is differentiable with reference to the network coefficients $\theta_q$ and $\theta_s$. Specifically, we have
\begin{equation}
    \frac{\partial G}{\partial \theta_q} = \sum_i \frac{\partial G}{\partial \mathbf{v}^d_i} 
    \left( \frac{\partial \mathbf{v}^d_i}{\partial \mathbf{q}} \frac{d \mathbf{q}}{d \theta_q}  \right) 
\; \mathrm{and} \;
\frac{\partial G}{\partial \theta_s} = \sum_i  \frac{\partial G}{\partial \mathbf{v}^d_i} 
    \left( \frac{\partial \mathbf{v}^d_i}{\partial \mathbf{s}} \frac{d \mathbf{s}}{d \theta_s}  \right), 
\end{equation}
among which ${d \mathbf{q}}/{d \theta_q}$ and  ${d \mathbf{s}}/{d \theta_s}$ can be directly obtained from the neural networks representing $\mathbf{q}(\mathbf{x})$ and $\mathbf{s}(\mathbf{x})$. By using the calculus of matrix, we can also have
\begin{equation}
    \frac{\partial \mathbf{\xi}}{\partial \mathbf{s}} =
    (\mathbf{A}\mathbf{\xi}^{0})^{T} \otimes
    ((\mathbf{A}^{T}\mathbf{A}+\mathbf{\gamma\mathbf{I}})^{-1}\mathbf{A}^{T}\mathbf{R})
\end{equation}
\begin{equation}
    \frac{\partial \mathbf{\xi}}{\partial \mathbf{q}} 
    = \frac{\partial \mathbf{\xi}}{\partial \mathbf{R}}
    \frac{\partial \mathbf{R}}{\partial \mathbf{q}} 
    = (\mathbf{s}\mathbf{A}\mathbf{\xi}^{0})^{T} \otimes((\mathbf{A}^{T}\mathbf{A}+\gamma\mathbf{I})^{-1}\mathbf{A}^{T})\frac{\partial \mathbf{R}}{\partial \mathbf{q}}
\end{equation}
where $\otimes$ denotes the Kronecker product and $\mathbf{R}$ is the rotation matrix form of $\mathbf{q}$. ${\partial \mathbf{v}^d_i}/{\partial \mathbf{s}}$ and ${\partial \mathbf{v}^d_i}/{\partial \mathbf{q}}$ can then be obtained from ${\partial \mathbf{\xi}}/{\partial \mathbf{s}}$ and ${\partial \mathbf{\xi}}/{\partial \mathbf{q}}$ from the corresponding location of $\mathbf{v}^d$ in $\xi$.

By the chain rule, the total derivatives from the loss function $\mathcal{L}$ to the network coefficients $\theta_q$ and $\theta_s$ are given as follows
\begin{equation}
 \frac{d \mathcal{L}}{d \theta_q} = \frac{\partial \mathcal{L}}{\partial G} \left( \sum_i \frac{\partial G}{\partial \mathbf{v}^d_i} 
    \left( \frac{\partial \mathbf{v}^d_i}{\partial \mathbf{q}} \frac{d \mathbf{q}}{d \theta_q}  \right) \right)
            + \frac{\partial \mathcal{L}}{\partial \mathbf{q}}\frac{d \mathbf{q}}{d \theta_q},
\end{equation}
\begin{equation}
\frac{d \mathcal{L}}{d \theta_s} = \frac{\partial \mathcal{L}}{\partial G} \left( \sum_i \frac{\partial G}{\partial \mathbf{v}^d_i} 
    \left( \frac{\partial \mathbf{v}^d_i}{\partial \mathbf{s}} \frac{d \mathbf{s}}{d \theta_s}  \right) \right)
            + \frac{\partial \mathcal{L}}{\partial \mathbf{s}}\frac{d \mathbf{s}}{d \theta_{s}}.
\end{equation}
The backpropagation of our NN-based computational pipeline can be realized. In practical implementation, the total derivatives can be computed by automatic differentiation \cite{paszke2019pytorch}. 

\subsection{Network Architecture and Constrained Optimization}\label{subsecNNArchOptm}
An architecture of \textit{Sinusoidal Representation Network} (SIREN) is adopted for $\mathbf{q}(\mathbf{x})$ and $\mathbf{s}(\mathbf{x})$ that contains $10$ hidden layers, each of which is formed by $512$ neurons with the periodic activation functions. The hyperparameters of this network are chosen by the strategy introduced in \cite{sitzmann2020implicit}. The input of the networks is $\mathbf{x} \in \mathbb{R}^3$, and the output is a four-dimensional vector for $\mathbf{q}(\mathbf{x}) \in \mathbb{R}^4$ and a three-dimensional vector as scaling ratios for $\mathbf{s}(\mathbf{x}) \in \mathbb{R}^3$. 
 
When computing the network coefficients according to the total loss, the Adam optimizer \cite{kingma2014adam} is employed due to its efficacy in handling large-scale data and complex computation. The initial learning rate is set to 1.0e-3, with a minimum learning rate threshold established at 1.0e-6. To adaptively adjust the learning rate during the optimization process, we utilize the `ReduceLROnPlateau` learning rate scheduler that can dynamically alter the learning rate based on the optimization progress.

To implement the optimization with hard constraints in our NN-based computational pipeline, we adopt the DC3 framework as introduced in \cite{donti2021dc}. Specifically, we implement a gradient-based correction step before the loss calculation and back-propagation. This involves taking gradient steps in the parameter space $(\theta_q, \theta_s)$ towards the feasible region with $\mathcal{L}_{CA} = 0$, thereby aligning solutions closer to the feasible region.

\begin{table*}[t]
    \caption{Computational statistics for Neural Slicer}
    \centering\label{tabCompStatistic}
    \small
    \begin{tabular}{r||ccccr||rrr|r||r}
    \hline 
        Model & Fig. & Objectives & \makecell[c]{Model \\ Genus} & \makecell[c]{Cage \\ Genus}  & \makecell[c]{\# of Cage \\ Elements} & \makecell[r]{Pre-processing \\ Time$^\dag$ (sec.)} & \makecell[r]{Optimization \\ Time (sec.)} & \makecell[r]{Post-processing\\ Time (sec.)} & \makecell[r]{Slicer Comp.\\Time (sec.)}& \makecell[r]{Toolpath Gen.\\Time (sec.)}\\
    \hline 
        Bunny Head & \ref{fig:Teaser} & SF+SR & $\textbf{22}$  & 0 & $23,139$ & $93.5$ ($60.1$) & $553.2$ & $124.1$ & $677.3$ & $52.7$\\
        Yoga & \ref{fig:overview} & SF+SR & 2 & 1 & $6,279$ & $55.1$ ($34.1$) & $312.3$ & $42.5$ & $354.8$ & $20.6$\\
        Shelf& \ref{fig:shelf_result} & SF+SR & $\textbf{30}$ & 26 & $42,150$ & $72.4$ ($52.3$) & $647.5$ & $104.1$ & $65.1$ & $65.1$\\
        Ring & \ref{fig:result-ring} & SF & 1  & 1 & $5,264$ & $33.4$ & $73.4$ & $15.4$ &  $88.8$ & $2.3$\\
        Tubes & \ref{fig:result-3rings} & SF & 5 & 4 & $16,551$ & $21.4$ & $153.0$ & $45.7$ & $3.5$ & $49.2$ \\
        Spiral Fish & \ref{fig:controlled-experiment} & SF & 3 & 3 & $43,147$ & $25.3$ & $223.5$ & $62.0$ & $285.5$ & $29.4$\\
        Bridge & \ref{fig:result-bridge} & SR & $\textbf{21}$ & 21 & $47,520$ & $52.1$ ($30.6$) & $174.3$  & $54.8$ & $229.1$ &$35.8$\\
    \hline 
    \end{tabular}
    \begin{flushleft}
    $^\dag$~Including the computing time for voxel-based FEA as given in the brackets.
    \end{flushleft}
    \label{tab:CptStatistics}
\end{table*}

\subsection{Pre-processing: Sampling and Cage Generation}\label{subsecCageGeneration}
As discussed above, a tetrahedral mesh $\mathcal{C}$ caging the input model $\mathcal{M}$ is constructed in our approach as the intermediate representation to determine the mapping $\mathbf{\lambda}(\cdot)$ according to $\mathbf{q}(\mathbf{x})$ and $\mathbf{s}(\mathbf{x})$. This representation for numerical computation is selected because of its good balance between the computational efficiency and the fidelity of deformation, making it highly suitable for our NN-based slicing framework. Meanwhile, the loss function of support-free is evaluated on a set of surface sampling points $\mathcal{B}$. The methods for generating $\mathcal{B}$ and $\mathcal{C}$ are introduced below.
\begin{figure}[t]
\centering
    \includegraphics[width=\linewidth]{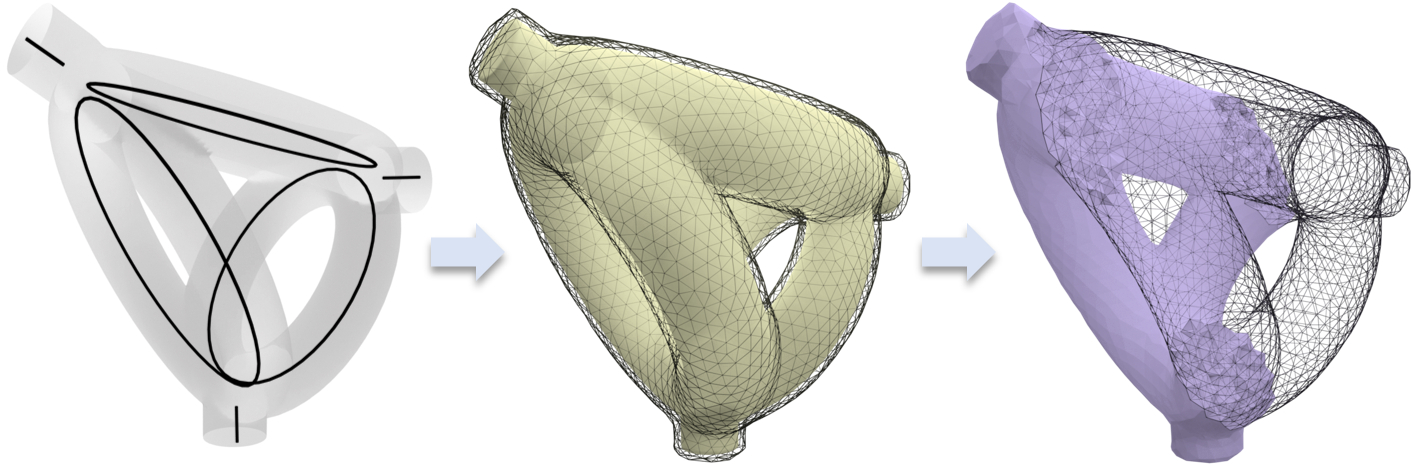}
    \put(-243,3){\small \color{black}(a)}
    \put(-165,3){\small \color{black}(b)}
    \put(-79,3){\small \color{black}(c)}
\caption{The illustration of cage generation: (a) the implicit surface $H(\mathbf{x})$ (gray) of the input model $\mathcal{M}$ in an abstract representation (e.g., as a convolution surface of the skeletons illustrated in black curves), (b) the polygonal mesh of $H(\mathbf{x})=0$ (yellow) and the caging surface mesh (black), (c) the tetrahedral mesh (purple) caging the input model.} 
\label{fig:CageGen}
\end{figure}

\begin{figure}
\centering
    \includegraphics[width=\linewidth]{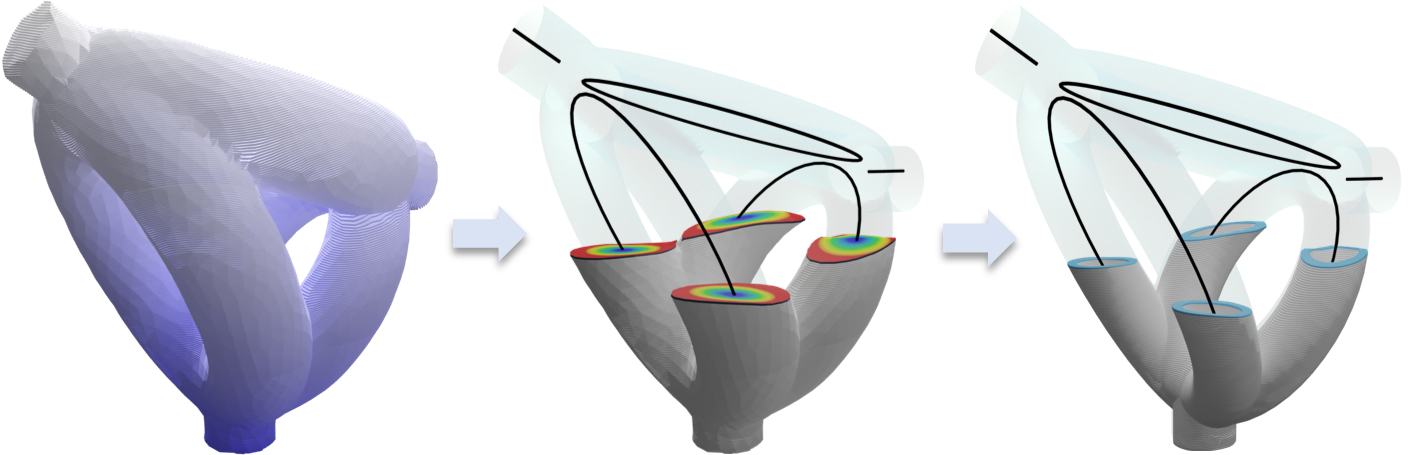}
    \put(-243,3){\small \color{black}(a)}
    \put(-162,3){\small \color{black}(b)}
    \put(-77,3){\small \color{black}(c)}
    \put(-200,70){\small \color{black}$G(\mathbf{x})$}
    \put(-40,70){\small \color{black}$H(\mathbf{x}) \leq 0$}
    \put(-153,26){\small \color{black}$\mathcal{G}_i$}
    \put(-68,26){\small \color{black}$\mathcal{P}_i$}
\caption{The illustration of slicing process to generate the curved layers: (a) a scalar field $G(\mathbf{x})$ that is optimized on the volumetric caging mesh $\mathcal{C}$, 
(b) the isosurfaces of $G(\mathbf{x})$ as polygonal surface meshes $\{\mathcal{G}_i \}$ are extracted from $\mathcal{C}$'s tetrahedra -- the colors on an isosurface visualize the function values of $H(\mathbf{x})$, and 
(c) the curved layers $\{\mathcal{P}_i \}$ are obtained by trimming $\{\mathcal{G}_i \}$ with the implicit solid $H(\mathbf{x}) \leq 0$.
}\label{fig:slicingOnCage}
\end{figure}
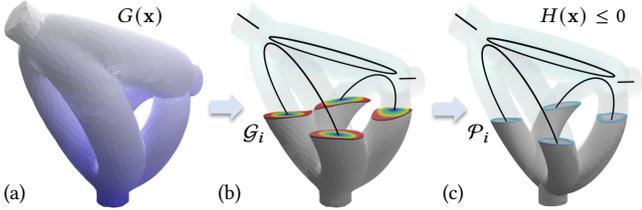 

After converting all different representations of the input model $\mathcal{M}$ into an implicit function $H(\mathbf{x})$ (see Fig.\ref{fig:CageGen}(a)), a polygonal mesh $\Tilde{\mathcal{H}}$ of its zero level-set surface $H(\mathbf{x})=0$ is extracted by the Marching Cubes algorithm~\cite{lorensen1998marching}. Surface points are sampled on $\Tilde{\mathcal{H}}$ and projected onto $H(\mathbf{x})=0$ by using the gradient $\nabla H(\mathbf{x})$. Sample points in $\mathcal{B}$ are obtained. 

A surface caging mesh can then be generated by using the Nested Cage approach~\cite{sacht2015nested}, which encloses all regions with $H(\mathbf{x}) \leq 0$ (see Fig.\ref{fig:CageGen}(b)). Lastly, the volumetric caging mesh $\mathcal{C}$ is constructed from the surface caging mesh by Tetgen~\cite{hang2015tetgen} (see Fig.\ref{fig:CageGen}(c)). The mapping and the curved layers are computed on this mesh $\mathcal{C}$.

\subsection{Post-processing: Slicing on Cage}\label{subsecSlicingOnCage}
The resultant curved layers are computed from the scalar field $G(\mathbf{x})$ that is defined on the volume caging mesh $\mathcal{C}$ (see Fig.\ref{fig:slicingOnCage}(a)). The isosurfaces of $G(\mathbf{x})$ are firstly extracted from the tetrahedra of $\mathcal{C}$ as piece-wise linear polygonal meshes $\{ \mathcal{G}_i \}$ (ref.~\cite{treece1999regularised}). The function values of $H(\mathbf{x})$ can then be evaluated at every point on a surface mesh $\mathcal{G}_i$ (see the values visualized as colors in Fig.\ref{fig:slicingOnCage}(b)). Lastly, the polygons of $\{\mathcal{G}_i\}$ are trimmed by the implicit solid $H(\mathbf{x}) \leq 0$ to form the curved layers $\{ \mathcal{P}_i \}$. Toolpaths are generated on each $\mathcal{P}_i$ and converted into trajectories of robot motion by the method of $S^3$-Slicer \cite{zhang2022s3}. 

\section{Result and Discussion}
Our computational pipeline is implemented in Python together with C++. PyTorch~\cite{paszke2019pytorch} is employed for constructing neural networks, automatic differentiation, and solving linear equation systems. PyVista ~\cite{sullivan2019pyvista} is employed for mesh processing. Time-consuming steps such as the point-to-surface distance query by~\cite{gottschalk1996obbtree}, the surface cage generation by~\cite{sacht2015nested}, and the toolpath generation by~\cite{zhang2022s3} are based on C++ implementation. The source code of our Neural Slicer will be released upon the acceptance of this paper.

\begin{figure*}[htbp]
    \centering
    \includegraphics[width=\linewidth]{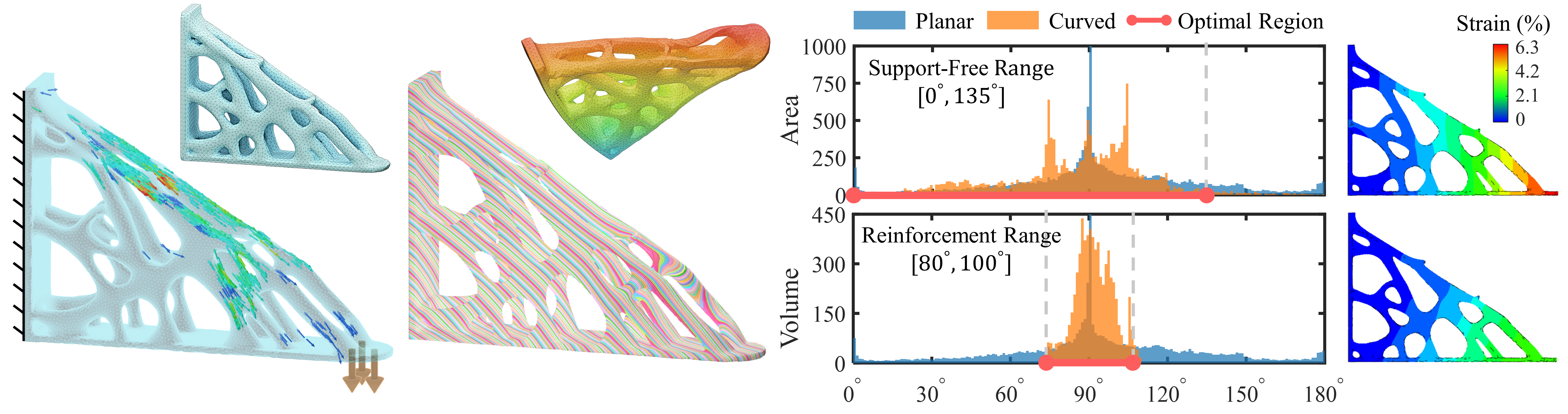}
    \put(-510,127){\small \color{black}(a)}
    \put(-380,127){\small \color{black}(b)}
    \put(-250,127){\small \color{black}(c)}
    \put(-70,127){\small \color{black}(d)}
    \put(-300,80){\small \color{black}Deformed}
    \put(-300,70){\small \color{black}Cage $\mathcal{C}^d$}
    \caption{The result generated by our Neural Slicer vs. the result by a planar slicer on the Shelf model: (a) the stress field under the given forces (shown as the arrows) and the model's cage used in the computation, (b) the curved layers (bottom) generated from the mapping determined by a deformed cage (top), (c) the histograms for evaluating the quality of results in terms of (top) the angles between LPDs and surface normals for the SF requirement and (bottom) the angles between LPDs and the maximal stresses for the SR requirement, (d) the results of FEA simulation by using anisotropic material orientations defined according to the LPDs for planar layers (top) and our curved layers (bottom).
    }
\label{fig:shelf_result}
\end{figure*}
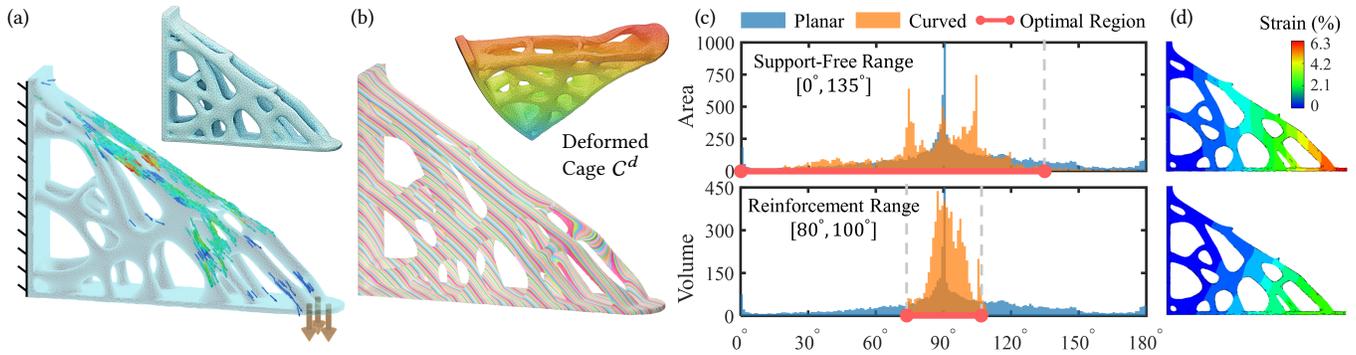

\subsection{Computational Experiments}
All the computational experiments are conducted on a desktop PC with an Intel(R) Core(TM) i5-12600K CPU (10 cores @3.6GHz), NVIDIA RTX 4080 GPU, and 32GB RAM, running Ubuntu 20.04 LTS (Focal Fossa). 

\subsubsection{Examples and computational statistics.}
We have tested our Neural Slicer on a variety of models with complicated geometry and topology. The first model is a Bunny Head model in hybrid representations (see Fig.\ref{fig:Teaser}(a)), including $15,635$ tetrahedra for solid (blue), $3,495$ triangles for shell (green), $41$ skeletons (thin black lines) for the struts as cylindrical solids and $36$ skeletons (bold black lines) for the tubular solids. The genus number of this input model is $g=22$. The implicit function $H(\textbf{x})$ of this model is computed with the help of the signed distance function and the convolution surfaces. Both the support-free (SF) and the strength reinforcement (SR) requirements are applied to this model. Curved layers can be successfully computed by our Neural Slicer (see Fig.\ref{fig:Teaser}(b)). The second example is the Yoga model with both the SF and SR objectives required as shown in Fig.\ref{fig:overview}, where the input is a mesh surface with $13,938$ triangles. The third example is a Shelf model generated by topology optimization. It is a model with a high genus number as $g=30$ and again both the SF and the SR requirements are applied. The resultant curved layers are shown in Fig.\ref{fig:shelf_result}. We can find that the curved layers follow the directions of maximal stresses very well. 

\begin{figure}
    \centering
    \includegraphics[width=\linewidth]{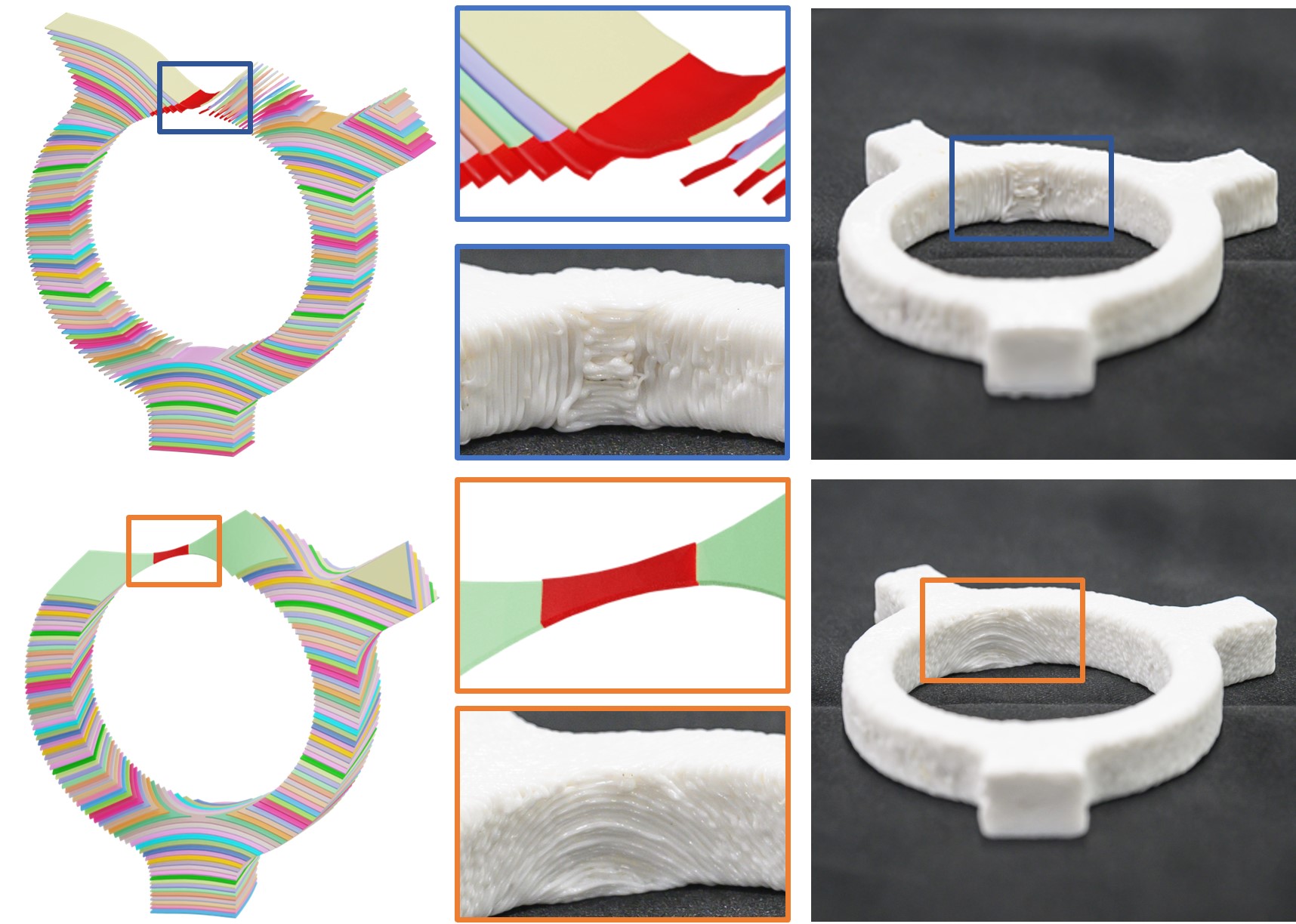}
    \put(-243,87){\small \color{black}(a)}
    \put(-243,2){\small \color{black}(b)}
\caption{Comparisons of the results generated by (a) $S^3$-Slicer and (b) our Neural Slicer for the Ring model.
}\label{fig:result-ring}
\end{figure}

We also tested our Neural Slicer on a few other models by applying only the SF requirement, including the Ring model (Fig.\ref{fig:result-ring}), the Tubes model (Fig.\ref{fig:result-3rings}) and the Spiral Fish model (Fig.\ref{fig:controlled-experiment}). The Ring model is selected to compare with the curved layers generated by $S^3$-Slicer. As illustrated in Fig.\ref{fig:Problem2_Def}, $S^3$-Slicer defined the SF objective in the deformed space. Layers that are self-supported in the deformed space can become overhangs in the model space caused by the mapping distortion. Differently, our Neural Slicer directly evaluates the SF loss in the model space so that the overhang can be better prevented (see the comparison shown in Fig.\ref{fig:result-ring}). The Tubes model is an implicit solid generated by convolution surfaces from the representation of $186$ line segments as skeletons. Our slicer can generate curved layers directly from the implicit solid. The Spiral Fish model is employed to demonstrate that our approach is robust to different initial guesses. Details will discussed later in Sec.~\ref{subsubsecInitGuess}. 

Lastly, we have tested this Neural Slicer on the Bridge example and compared our approach with the $S^3$-Slicer (see Fig.\ref{fig:result-bridge}). Only the SR requirement is applied in this example. Computational statistics of all examples are given in Table \ref{tab:CptStatistics}, where the computation of our Neural Slicer can be completed within 15 minutes. It can be observed that the genus number of a caging mesh can be different from an input model. In other words, we can use a caging mesh with simple topology as the computational domain to generate curved layers for a model with complicated topology (see the Bunny Head model and its cage in Fig.\ref{fig:Teaser} for an example). The toolpaths on the curved layers are generated by the method presented in \cite{zhang2022s3}, the computing time of which has also been reported in Table \ref{tab:CptStatistics}.

\subsubsection{Statistics in LPDs}
As both the SF and the SR requirements are defined according to the LPDs, we evaluate the quality of results generated by our Neural Slicer as the histograms in terms of i) the angles between LPDs and the surface normals at $H(\textbf{x})=0$ for the SF requirement and ii) the angles between LPDs and the maximal stresses for the SR requirement. The histograms are generated by LPDs (i.e., $ \nabla G(\mathbf{x}) / \| \nabla G(\mathbf{x}) \|$) evaluated at all the sample points in the set $\mathcal{B}$ for SF and the set $\mathcal{T}$ for SR. We have given these statistical results as histograms for the Bunny Head model in Fig.\ref{fig:Teaser} and the Shelf model in Fig.\ref{fig:shelf_result}. It can be observed from the histograms that both the SF and the SR requirements have been well achieved on the curved layers generated by our Neural Slicer. Despite this, supporting structures are still needed to help mount the Shelf model tightly on the platform during the printing process.

\begin{figure}[t]
    \centering
    \includegraphics[width=\linewidth]{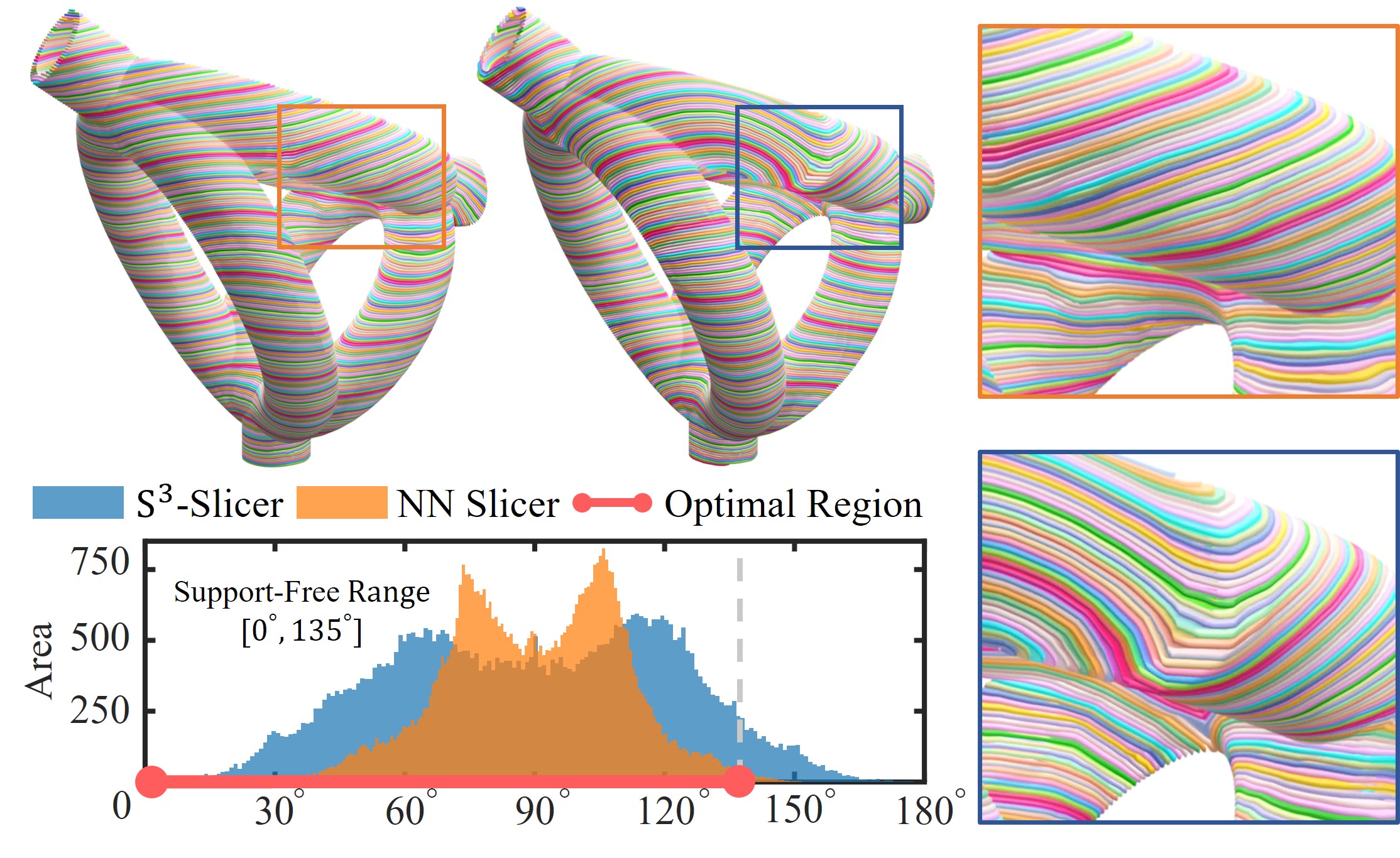}
    \put(-243,73){\small \color{black}(a)}
    \put(-163,73){\small \color{black}(b)}
    \put(-243,3){\small \color{black}(c)}
\caption{The results of the Tubes model that is generated by convolution surface from a skeleton representation (see Fig.\ref{fig:CageGen}(a)): the curved layers generated by (a) our method and (b) $S^3$-Slicer. (c) The histograms of the angles between LPDs and the surface normals are used to visualize the quality of the SF objective achieved.
}\label{fig:result-3rings}
\end{figure}

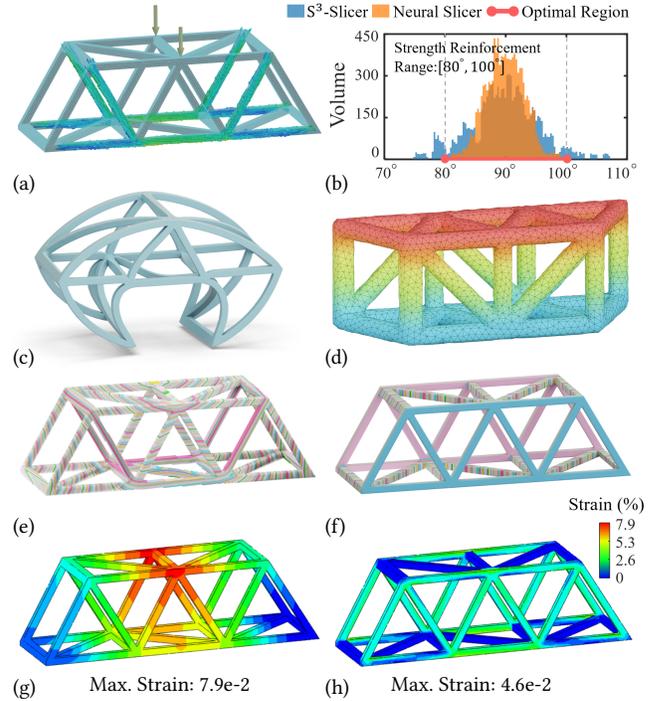
\begin{figure}[t]
    \centering
    \includegraphics[width=\linewidth]{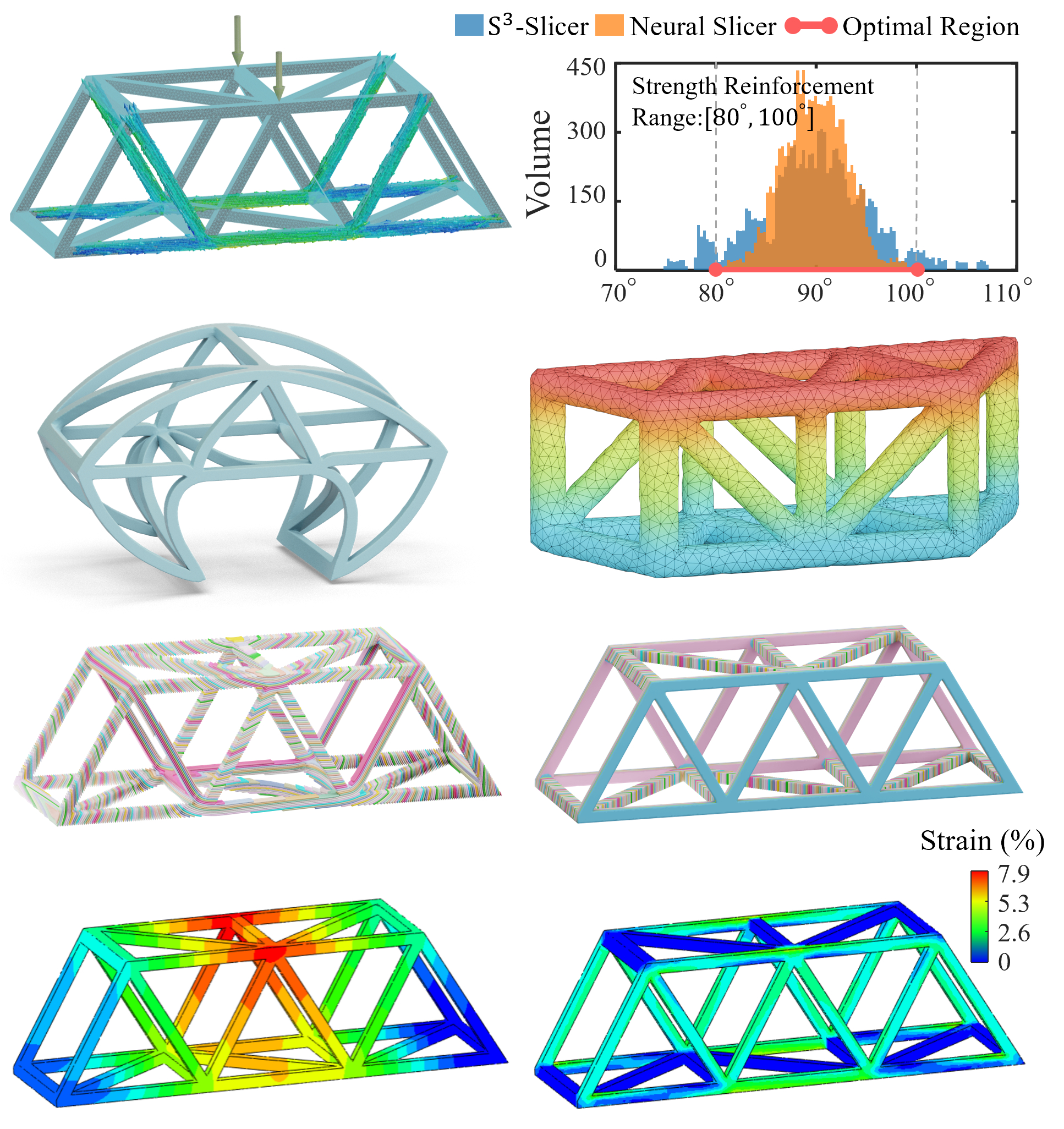}
    \put(-243,185){\small \color{black}(a)}
    \put(-125,185){\small \color{black}(b)}
    \put(-243,119){\small \color{black}(c)}
    \put(-125,119){\small \color{black}(d)}
    \put(-243,55){\small \color{black}(e)}
    \put(-125,55){\small \color{black}(f)}
    \put(-243,-6){\small \color{black}(g)}
    \put(-125,-6){\small \color{black}(h)}
    \put(-214,-5){\small \color{black}Max. Strain: 7.9e-2}
    \put(-100,-5){\small \color{black}Max. Strain: 4.6e-2}
\caption{Comparing our results with the result of $S^3$-Slicer on the Bridge model for the SR requirement: 
(a) the input model and the stress field under the given loading, 
(b) the histogram to visualize the results in terms of angles between LPDs and maximal principal stresses, 
(c \& e) the deformed shape obtained by $S^3$-Slicer and its resultant curved layers, (d \& f) the deformed caging mesh generated by our Neural Slicer and the corresponding curved layers. Our approach can find a very `smart' solution that has nearly planar layers that align with the distribution of maximal stresses better. To further verify the mechanical strength, the anisotropic FEA is conducted to generate strain distribution for the results of (g) $S^3$-Slicer and (h) our Neural Slicer.
 }\label{fig:result-bridge}
\end{figure}

The statistics for the curved layers generated by the $S^3$-slicer \cite{zhang2022s3} are also given on the Tubes model (Fig.\ref{fig:result-3rings}) and the Bridge model (Fig.\ref{fig:result-bridge}) for the purpose of comparison. It is found that our Neural Slicer can further reduce the remaining overhang region by $95\%$ on the Tubes model. The results are also visualized by the histogram shown in Fig.\ref{fig:result-3rings}(c). Note that the result of the Tubes model is computed by $S^3$-slicer on a very dense tetrahedral mesh with 430k elements while our Neural Slicer only employs a caging mesh with 16.5k elements.

When applying our Neural Slicer on the Bridge model for the SR requirement (see Fig.\ref{fig:result-bridge}), we can generate a result of curved layers that outperform those obtained from the $S^3$-slicer~\cite{zhang2022s3}. Instead of the highly curved layers, our slicer automatically generates a very `smart' solution with nearly planar layers (see Fig.\ref{fig:result-bridge}(d \& f)), which surprisingly aligns better with the maximal stresses -- see the histogram given in Fig.\ref{fig:result-bridge}(b). 

\subsubsection{Verification by FEA}\label{subsubsecFEA} 
To verify the mechanical strength of models to be fabricated by different layers, we conducted the FEA simulation with anisotropic material properties by assigning different Young’s modulus along different directions at the element level. Specifically, $Y_{1} = 3.5~\mathrm{GPa}$ is used as the strongest modulus and is assigned to the toolpath's tangential direction. The weakest modulus is assigned both to the surface normal direction of each layer and the third orthogonal direction as $Y_{2} = Y_{3} = 1.2~\mathrm{GPa}$. 

\begin{figure}[t]
    \centering
    \includegraphics[width=\linewidth]{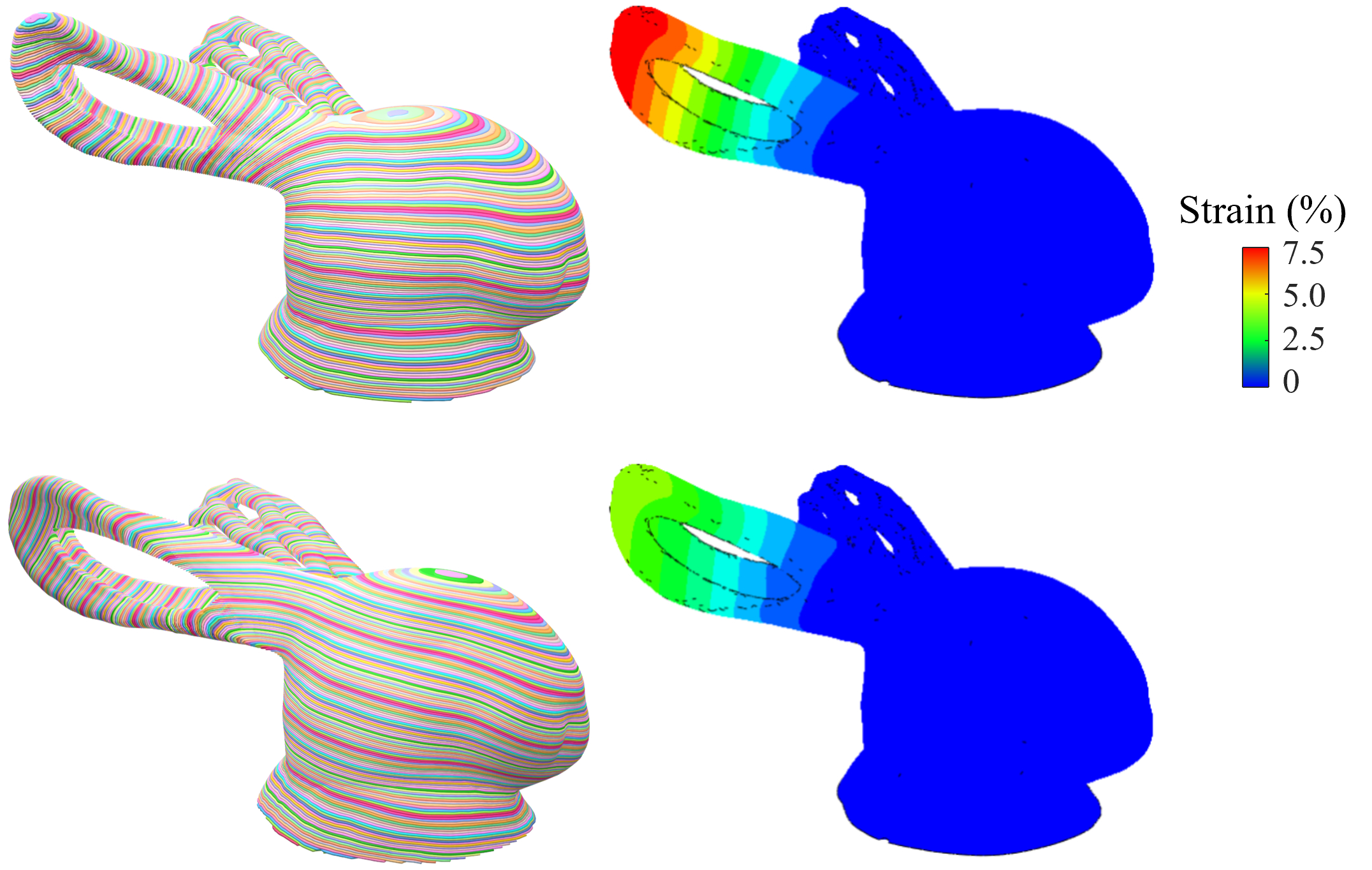}
    \put(-243,85){\small \color{black}(a)}
    \put(-132,85){\small \color{black}(b)}
    \put(-243,0){\small \color{black}(c)}
    \put(-132,0){\small \color{black}(d)}
    \put(-70,150){\small \color{black}Max. Strain: 3.8e-2}
    \put(-70,67){\small \color{black}Max. Strain: 2.4e-2}
\caption{The results of FEA simulation with anisotropic material properties on the Bunny Head model using curved layers generated by our Neural Slicer with (a \& b) only the SF requirement and (c \& d) the SF + SR requirements.
}\label{fig:bunnyhead-FEA}
\end{figure}

FEA results of the Shelf model have been given in Fig.\ref{fig:shelf_result}(d), where the model with curved layers generated by our Neural Slicer has reduced the maximal strain by  $43.3\%$ compared to the model using planar layers. For the FEA simulation conducted on the Bridge model, the Neural Slicer can reduce the maximal strain by $40.5\%$ than the $S^3$-Slicer (see Fig.\ref{fig:result-bridge}(g \& h)). In Fig.\ref{fig:bunnyhead-FEA}, we compare the FEA results on the Bunny Head model for the results with only the SF requirement vs. with both the SF and the SR requirements. Adding the SR requirement can reduce the maximal strain by $36.8\%$.

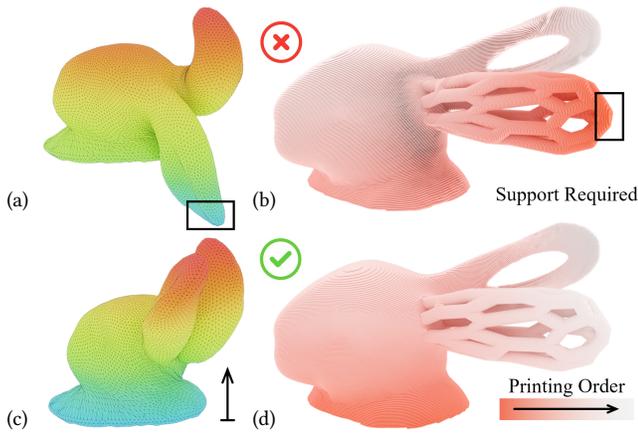
\begin{figure}[t]
    \centering
    \includegraphics[width=\linewidth]{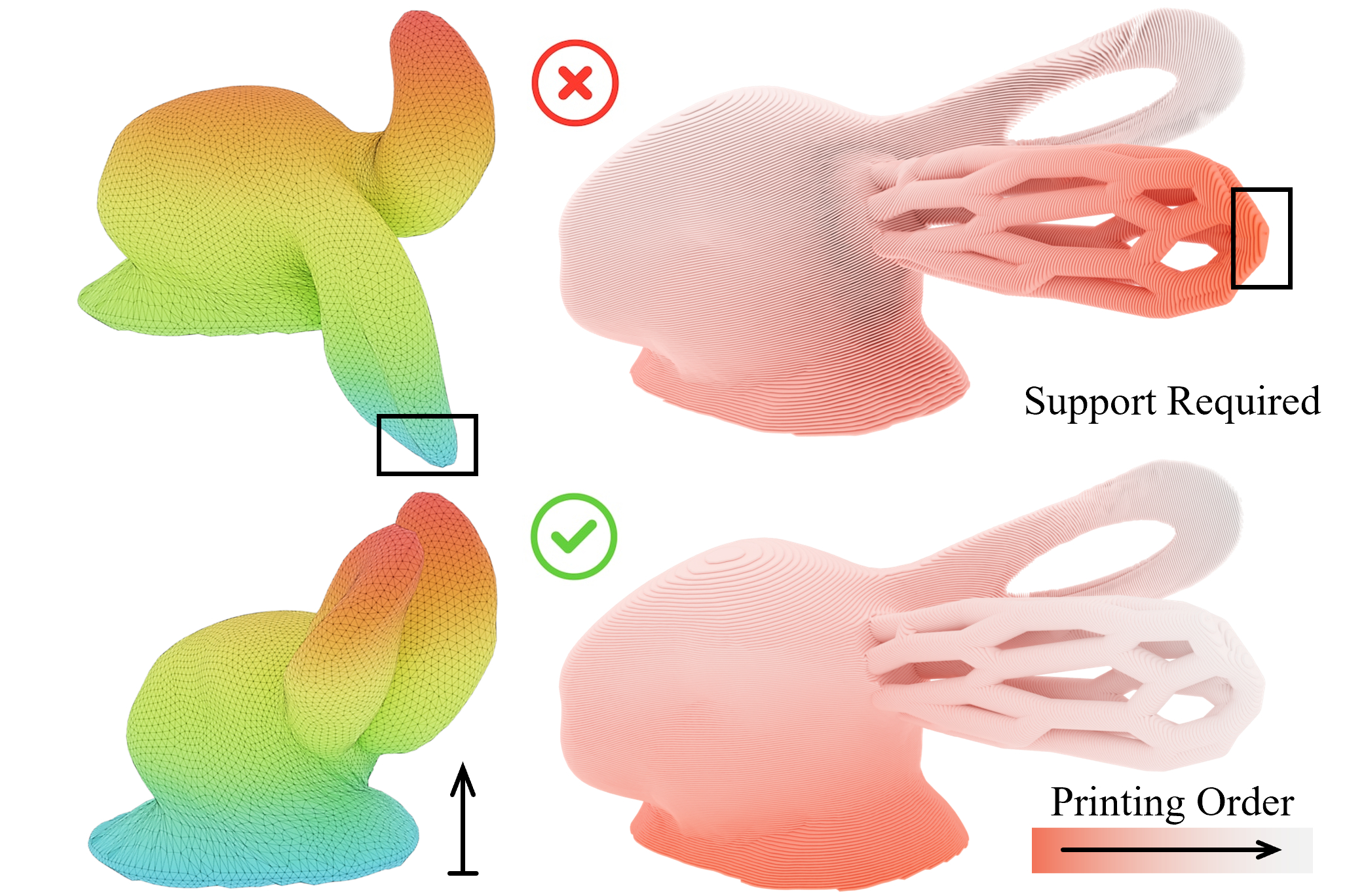}
    \put(-243,85){\small \color{black}(a)}
    \put(-150,85){\small \color{black}(b)}
    \put(-243,2){\small \color{black}(c)}
    \put(-150,2){\small \color{black}(d)}
    \caption{Ablation study of the point overhang loss $\mathcal{L}_{PO}$ on the Bunny Head model: (a \& c) illustrate the deformed cages $\mathcal{C}^d$ without / with $\mathcal{L}_{PO}$, (b \& d) show the printing order of slices by colors (from pink to gray). Printing sequence that starts from the ear region is not manufacturable as the materials will be deposed onto the `air'.
}\label{fig:ablation-experiments}
\end{figure}

\subsubsection{Ablation Study for Point Overhang Loss}
To demonstrate the effectiveness of the point overhang loss $\mathcal{L}_{PO}$ (Eq.(\ref{eqPOLoss})), we conducted an ablation study on the Bunny Head model imposing both the SF and the SR requirements. The only difference is whether the point overhang loss is omitted. When skipping the point overhang loss, the optimization will lead to a deformed caging mesh $\mathcal{C}^d$ as shown in Fig.\ref{fig:ablation-experiments}(a). Compared to the result with the full set of losses (as Fig.\ref{fig:ablation-experiments}(c)), this omission leads to an overhang at the tip of the right ear. As a result, the printing sequence determined according to $\mathcal{C}^d$ in Fig.\ref{fig:ablation-experiments}(b) will require additional support at the tip of the right ear. Differently, the printing sequence given in Fig.\ref{fig:ablation-experiments}(d) can be realized in a completely support-free way.  

\subsubsection{Robustness to Initial Guess}\label{subsubsecInitGuess}
One major issue of previous approaches based on nonlinear optimization is how to obtain good initial guesses for LPDs. As can be found in the example of the Tubes model shown in Fig.\ref{fig:result-3rings}, the $S^3$-Slicer computes results by using the distance field generated by the heat method \cite{Crane_17Communication_ACM} as an initial guess. Apparently, the resultant curved layers are less smoother than those generated by our Neural Slicer (see the zoom views of Fig.\ref{fig:result-3rings}). Another example is the Bridge model as shown in Fig.\ref{fig:result-bridge}. When posing the input model into an orientation as Fig.\ref{fig:result-bridge}(d) and using the height field as the initial guess for LPDs, the $S^3$-Slicer can obtain a result similar to our Neural Slicer. However, a `good' initial guess is never easy to be obtained in general. 

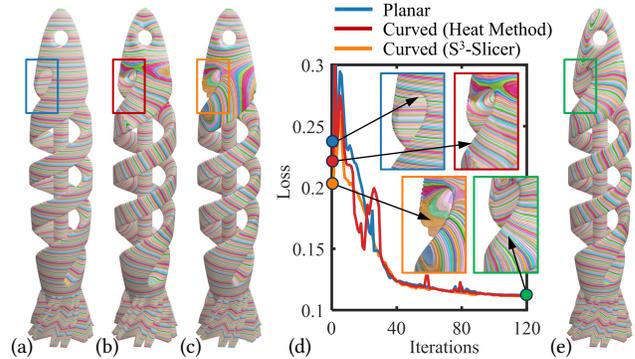
\begin{figure}[t]
    \centering
    \includegraphics[width=\linewidth]{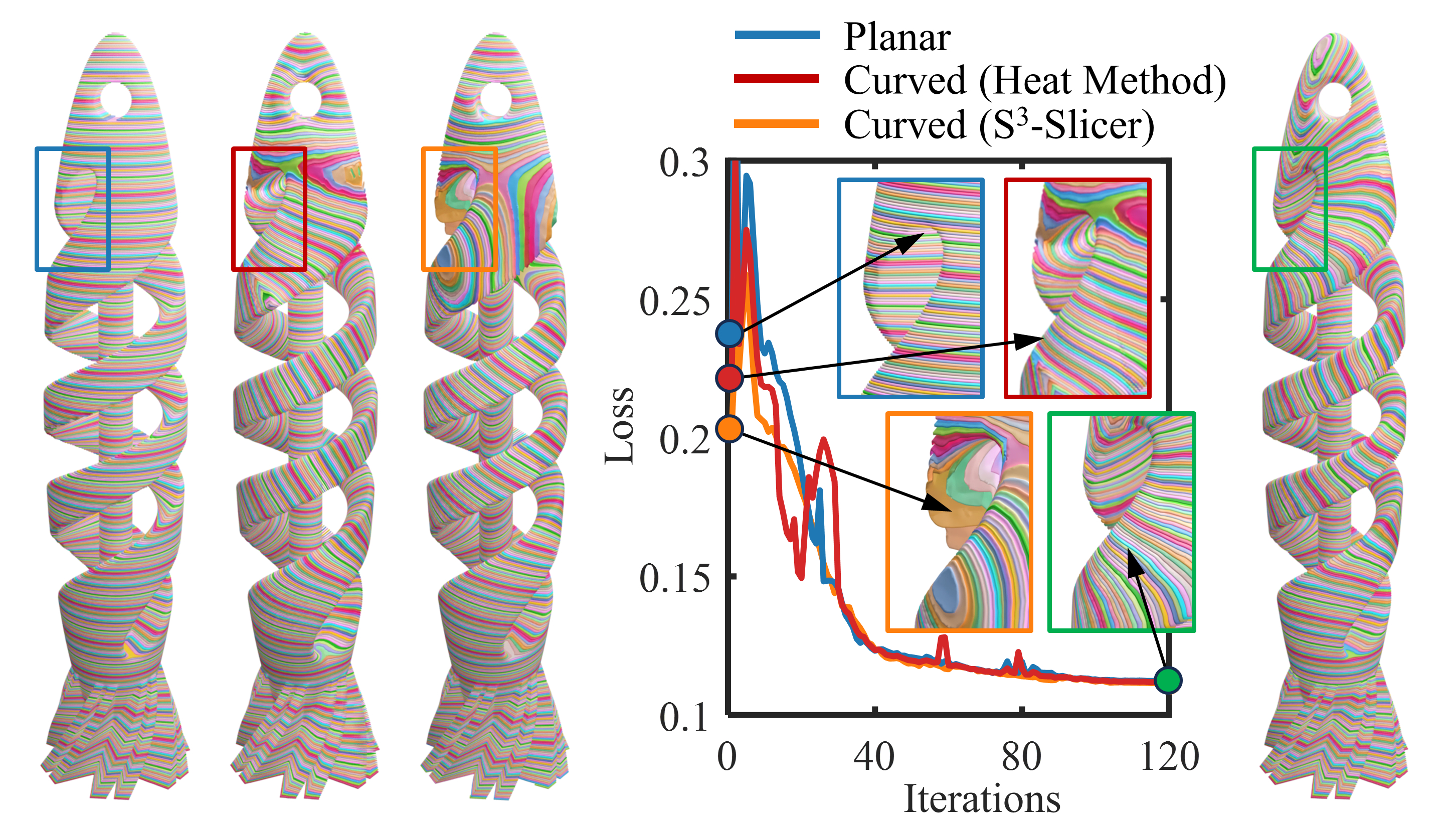}
    \put(-243,2){\small \color{black}(a)}
    \put(-211,2){\small \color{black}(b)}
    \put(-179,2){\small \color{black}(c)}
    \put(-138,2){\small \color{black}(d)}
    \put(-39,2){\small \color{black}(e)}
    \caption{Study conducted on a Spiral-Fish model by using different initial guesses for our Neural Slicer: (a) planar layers as a height field, (b) curved layers from a field generated by the heat method~\cite{Crane_17Communication_ACM} and (c) curved layers as the result of $S^3$-Slicer~\cite{zhang2022s3}. The optimization as a network learning process (see the learning curves in (d)) can always converge to the result with curved layers as shown in (e).
}
    \label{fig:controlled-experiment}
\end{figure}

Differently, our Neural Slicer is robust to the initial guess of the mapping $\lambda(\cdot)$. A study has been conducted on the Spiral Fish model as shown in Fig.\ref{fig:controlled-experiment} to demonstrate this advantage of our approach. The computing processes starting from different initial guesses are tested, which include:
\begin{enumerate}
\item Height Field -- We first test the initial guess as planar layers, which is in fact an identity mapping for $\lambda(\cdot)$ that can be easily realized by setting $\mathbf{q(\mathbf{x})}$ and $\mathbf{s(\mathbf{x})}$ with neither rotation nor scaling (see Fig.\ref{fig:controlled-experiment}(a)). 

\item Heat Transfer -- The other initial field is obtained by using the heat method \cite{Crane_17Communication_ACM} to generate an approximated geodesic distance field to the bottom of the model (see Fig.\ref{fig:controlled-experiment}(b)). The network coefficients $\theta_q$ and $\theta_s$ for $\mathbf{q(\mathbf{x})}$ and $\mathbf{s(\mathbf{x})}$ are pre-trained to make $G(\textbf{x})$ fit this field by minimizing the mean squared error. 

\item $S^3$-Slicer -- The scalar field obtained from the $S^3$-Slicer (see Fig.\ref{fig:controlled-experiment}(c)) is employed to pre-train the initial network coefficients $\theta_q$ and $\theta_s$ for $\mathbf{q(\mathbf{x})}$ and $\mathbf{s(\mathbf{x})}$.
\end{enumerate}
When applying our Neural Slicer to compute curved layers by using the above different initial guesses, the learning curves of the NN optimizer are as shown in Fig.\ref{fig:controlled-experiment}(d). All converge to the same result of curved layers as shown in Fig.\ref{fig:controlled-experiment}(e). The area of overhang on our result can be further reduced by $94.2\%$ w.r.t. the result from $S^3$-Slicer. This study demonstrates the robustness of our framework with different initial guesses.

\subsection{Physical Experiments}\label{subsecPhysicalTest}
The curved layers generated by our framework are tested to fabricate models on a multi-axis 3D printing hardware system as shown in Fig.\ref{fig:Teaser}(d), which is composed of an ABB IRB 2600 robotic arm with 6-DOFs and an ABB A250 positioner with 2-DOFs. The repeatability of this robotic system is $\pm0.05$mm and the TCP/IP protocol is used for the communication between the extrusion system, the robotic arms, and the PC. With the kinematic redundancy provided by this system, materials can be deposed onto the in-process model along arbitrary directions in the model's space while still controlling the physical orientation and the speed / acceleration of the printer head. The extruder system is controlled by a Duet3D board and the nozzle dimension is $1.0$mm. Polylactic Acid (PLA) and Polyvinyl Alcohol (PVA) filaments with 1.75mm diameter are used in our physical fabrication, where all models are printed by PLA filaments. PVA is only used to print the support structures for the Bridge model. With the help of speed control, layers with thickness in the range of $[0.4\mathrm{mm}, 1.0\mathrm{mm}]$ can be reliably produced by this hardware system.
\begin{figure}[t]
\centering
\includegraphics[width=\linewidth]{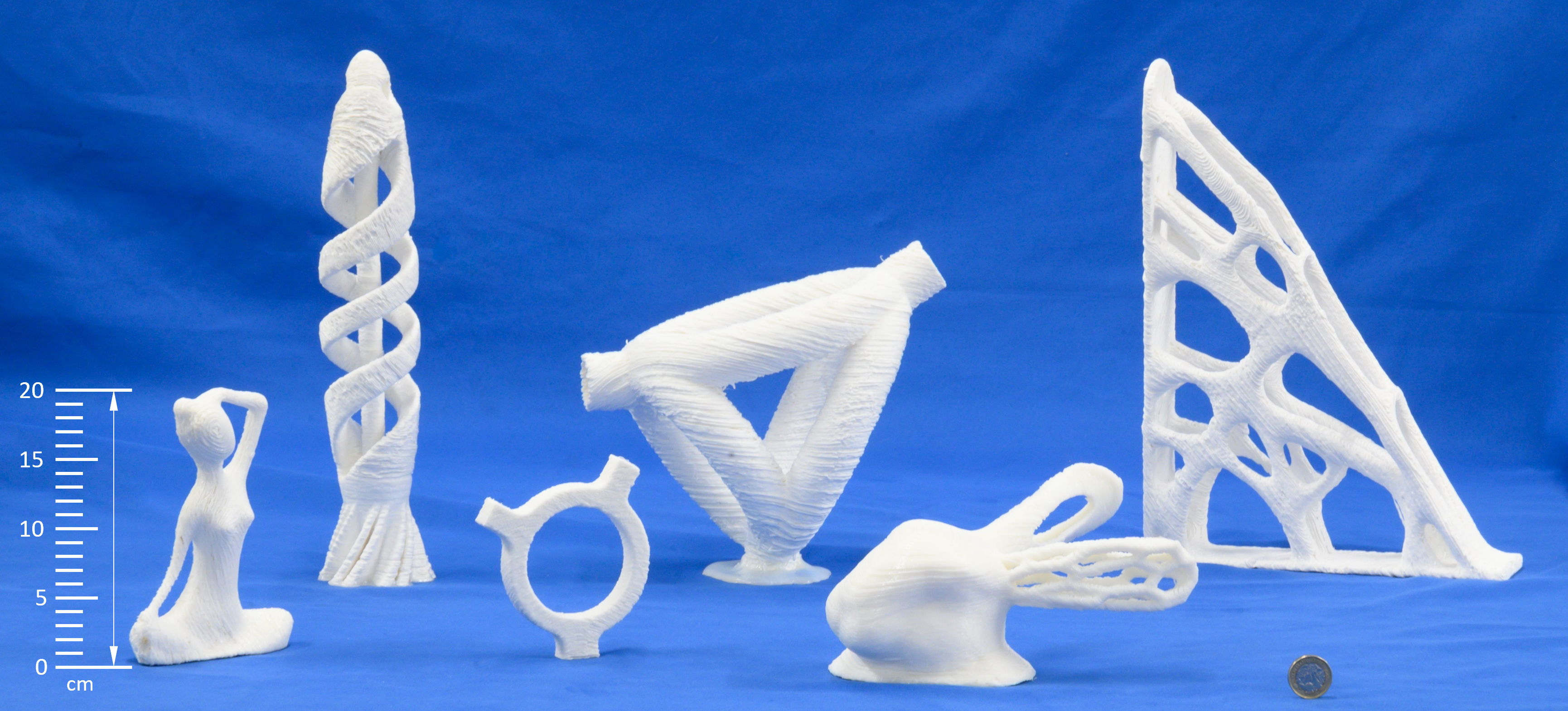}
\caption{The results of all models that are fabricated using the support-free curved layers generated by our Neural Slicer.}\label{fig:allPhysicalResult}
\end{figure}

\begin{table}[t]
\caption{Statistics of physical fabrication.}
\centering\label{tab:FabricationStatistic}
\small
\begin{tabular}{l||c|r|r || c|c|r }
\hline
 & \multicolumn{3}{c||}{Curved Layers$^\dag$} & \multicolumn{3}{c}{Planar Layers$^\dag$}\\
\cline{2-7}
Model   &  \#Layer & Weight & Time & \#Layer & Weight$^\ddag$ & Time  \\
\hline \hline
Bunny Hd. &  $250$  &  $512.9$g  &  21.3h  & $243$  &  $659.4$g   &  $24.6$h  \\
Yoga   &  150  &  $140.8$g   &  18.7h  & /  &  /   &  /  \\
Shelf   &  320  &  $521.0$g  &  39.3h  & /  &  /   &  /  \\
Ring   &  $99$  &  $52.3$g   &  2.4h  & $87$  &  $100.1$g   &  $4.5$h  \\
Tubes     &  $150$  &  $324.7$g   &  22.3h  & $132$  &  $514.8$g   &  $20.9$h \\
Spiral Fish  &  $500$  &  $326.6$g   &  24.8h  & $340$  &  $445.5$g   &  $19.5$h \\
\hline
Bridge (Neu)   &  350  &  $769.1$g   &  26.2h  & /  &  /   &  /  \\
Bridge ($S^3$)   &  600  &  $921.1$g   &  44.8h  & /  &  /   &  /  \\
\hline
\end{tabular}
\begin{flushleft}
$^\dag$~The thicknesses for curved layers are in the range of [$0.4$mm, $1.0$mm], and the thickness for planar layers is $0.8$mm. \\
$^\ddag$~The weight of a model fabricated by planar layers includes the weight of its supporting structures.
\end{flushleft}
\end{table}

\begin{figure}[t]
\centering
\includegraphics[width=\linewidth]{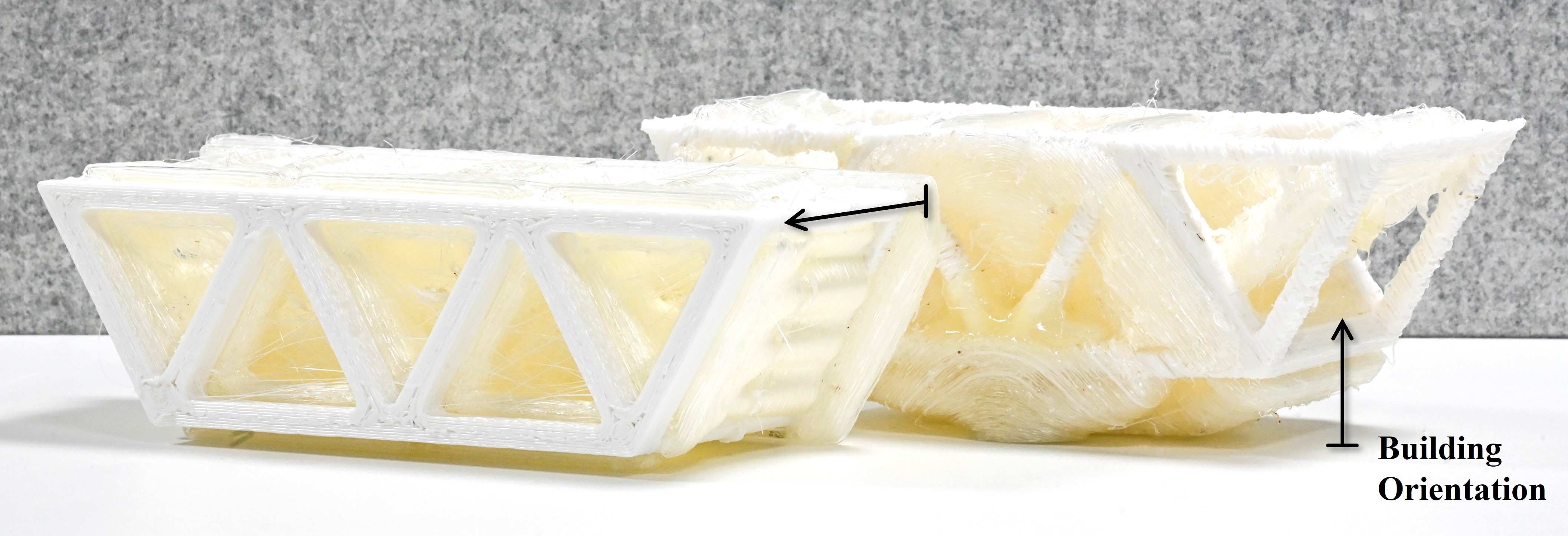}
\put(-235,3){\small \color{black}(a)}
\put(-116,3){\small \color{black}(b)}
\caption{Fabrication results of the Bridge model using (a) the layers generated by our Neural Slicer and (b) the layers generated by the $S^3$-Slicer, where the building orientations have been specified by the arrows.
}\label{fig:result-BridgeFabRes}
\end{figure}

\begin{figure}[t]
\centering
\includegraphics[width=\linewidth]{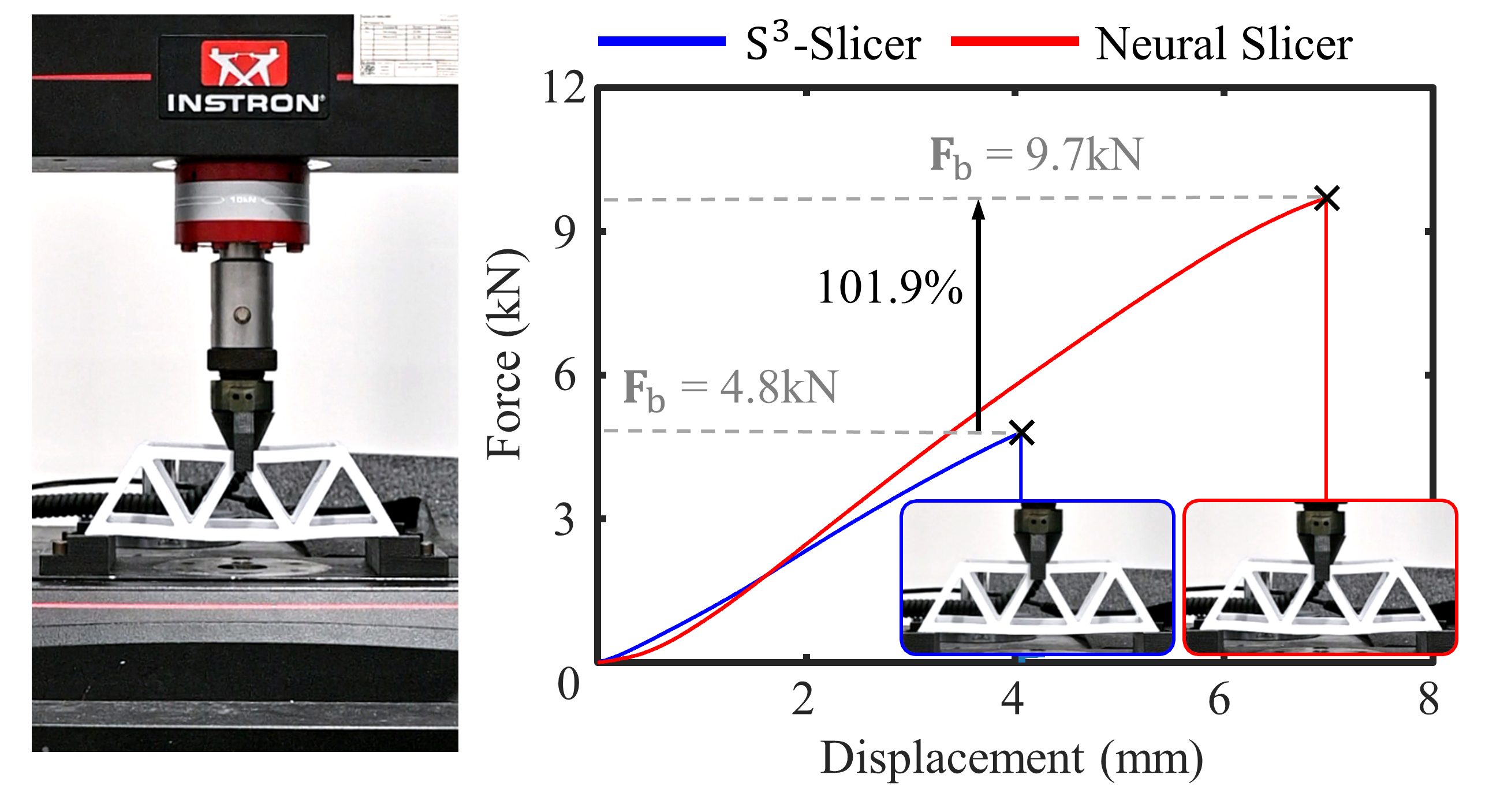}
\caption{Results of the Bridge model under 3-Point bending test, where the force is applied at the top-middle of the Bridge and the bottom of the model is fixed. Force-displacement curves are generated to study the mechanical strength of models fabricated from different curved layers. The breaking force has been doubled ($ \uparrow 101.9\%$) compared to the result of $S^3$-Slicer~\cite{zhang2022s3}.
}\label{fig:result-bridgeMechTest}
\end{figure}

\begin{figure}[t]
\centering
\includegraphics[width=\linewidth]{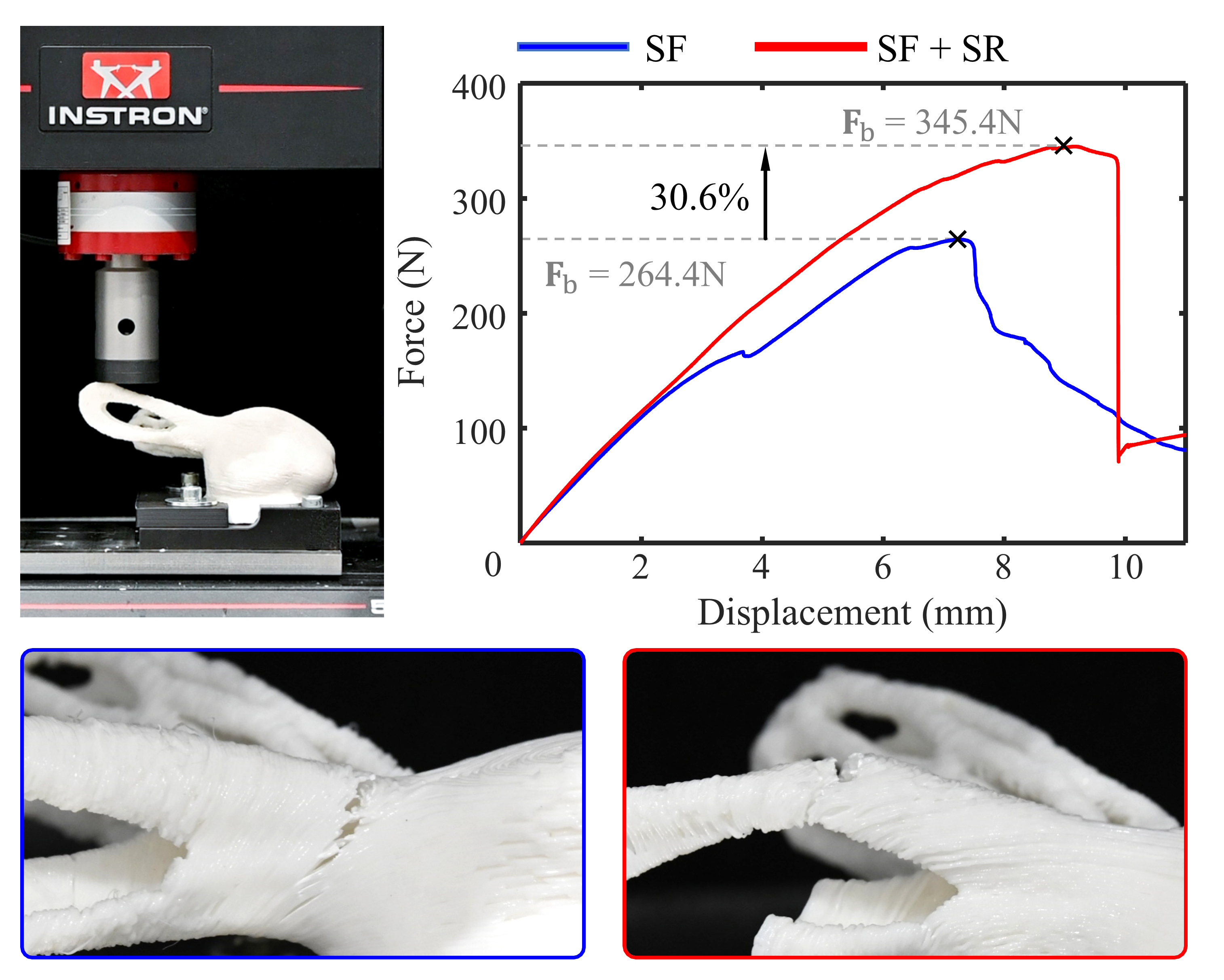}
\caption{The results of compression tests taken on two specimens of the Bunny Head model, where the force is applied on the right ear of the model. The specimens are fabricated from layers with only the SF requirement and the SF + SR requirements -- both are generated by our Neural Slicer. After imposing the SR requirement, the model's breaking force can be increased by $30.6\%$.
}\label{fig:result-bunnyHeadMechTest}
\end{figure}

The models fabricated by our system using support-free curved layers are shown in Fig.\ref{fig:allPhysicalResult}. The statistics of physical fabrication are given in Table~\ref{tab:FabricationStatistic}. For the purpose of comparison, a few models are also fabricated by planar layers, where additional supporting structures need to be printed therefore the models are heavier. To have a larger search space, curved layers of the Bridge model are computed by only imposing the SR requirement. Two Bridge models are fabricated using layers generated by the $S^3$-Slicer and our Neural Slicer (see Fig.\ref{fig:result-BridgeFabRes}).

Mechanical tests have been conducted to test the strength of 3D printed models on an INSTRON tensile machine 5960 with a capacity of 10kN. First of all, the 3-point bending tests are conducted on two Bridge models shown in Fig.\ref{fig:result-BridgeFabRes} after removing the PVA supports. Force-displacement curves are measured during the bending tests (see Fig.\ref{fig:result-bridgeMechTest}). An improvement of $101.9\%$ in the breaking force can be observed on the specimen fabricated from the layers generated by our Neural Slicer, which is consistent with the simulation results as given in Fig.\ref{fig:result-bridge}. We also tested the mechanical strength of the 3D printed Bunny Head model on the same tensile machine. The bottom of the model is fixed by a specially designed fixture and the compressing force is applied on the right ear of the model (see Fig.\ref{fig:result-bunnyHeadMechTest}). Two specimens are tested in our experiment. One is fabricated from the layers generated by imposing only the SF requirement while the other is from the layered with both the SF and the SR requirements. From the force-displacement curves, we can observe a $30.6\%$ increase of the breaking force on the specimen with SR requirement. The broken region is changed from the root of the ear to the hole of the ear (see the zoom-views in Fig.\ref{fig:result-bunnyHeadMechTest}). The results of mechanical tests are also consistent with the simulation results given in Fig.\ref{fig:bunnyhead-FEA}, where the maximal strain on the model with the SF + SR requirements is reduced by $36.5\%$ compared to the one with only the SF requirement. 

\subsection{Discussion}
\begin{figure}[t]
\centering
\includegraphics[width=\linewidth]{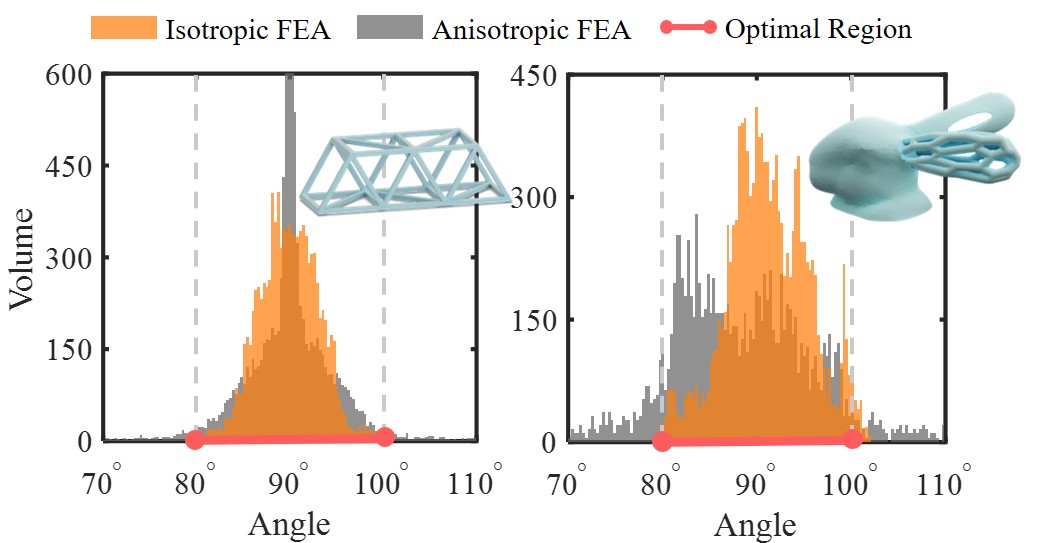}
\caption{Statistics of the angles between LPDs and the directions of maximal stresses obtained by isotropic FEA vs. anisotropic FEA.
}\label{fig:taumax_angleChange}
\end{figure}

One major limitation of our current approach is that the stress field employed to define the strength reinforcement loss $\mathcal{L}_{SR}$ is obtained from FEA using isotropic material properties (i.e., $\mathbf{\tau}_{\max}$ is given and unchanged during the optimization). This simplification ignores the change of $\mathbf{\tau}_{\max}$ caused by the anisotropic material properties introduced by curved layers. We study this influence by studying the change of angles between LPDs and the maximal stress directions when using anisotropic FEA instead of the input distribution obtained from isotropic FEA. The anisotropic FEA in this study employs the same parameters as those used in Sec.~\ref{subsubsecFEA}. The statistics of angle changes on two models have been given in Fig.\ref{fig:taumax_angleChange}. It can be found that this change indeed has a certain level of influence but is not very significant. Moreover, considering the change of $\mathbf{\tau}_{\max}$ in curved slicing needs to incorporate the anisotropic FEA in the loop of optimization, which will further increase the computational complexity. How to make the procedure of principal stress analysis differentiable is also a problem to be further explored. We plan to consider this in our future research.

\begin{figure}[t]
\centering
\includegraphics[width=\linewidth]{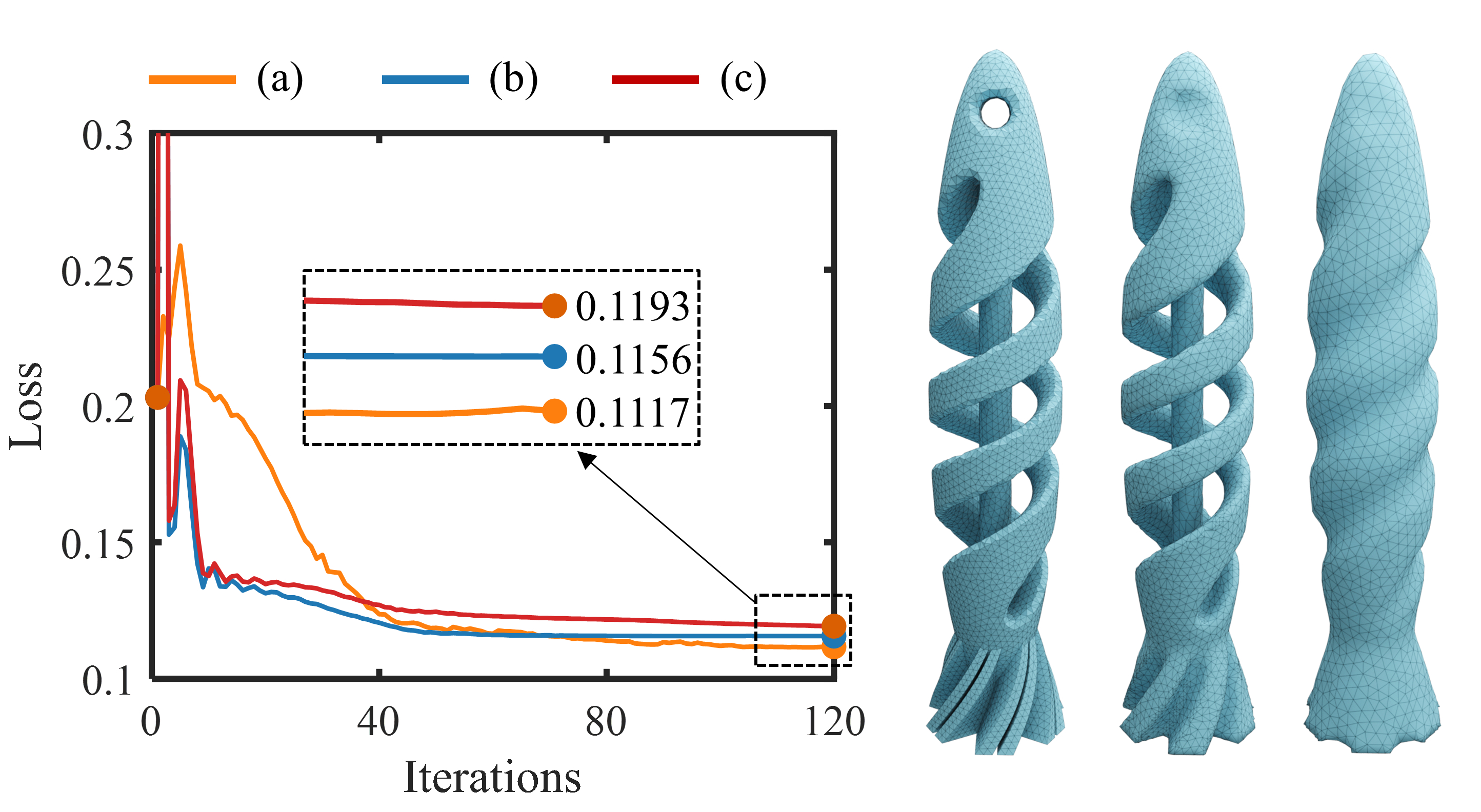}
    \put(-94,2){\small \color{black}(a)}
    \put(-61,2){\small \color{black}(b)}
    \put(-30,2){\small \color{black}(c)}
\caption{The learning curves for the Spiral Fish model by using the caging mesh in different resolutions -- i.e., with (a) $43,147$ elements ($g=3$), (b) $35,453$ ($g=2$) elements and (c) $16,153$ elements ($g=0$), where $g$ indicates the genus number of a caging mesh. To conduct a fair comparison, only $\mathcal{L}_{SF} + \mathcal{L}_{PO}$ are visualized in the learning curves. 
}\label{fig:spiralFish_CageResStudy}
\end{figure}

The second limitation is that the computation of mapping is based on an intermediate representation -- the caging mesh $\mathcal{C}$. We conducted the computational tests on the Spiral Fish model by using $\mathcal{C}$ in three different resolutions. As shown in Fig.\ref{fig:spiralFish_CageResStudy}, the computation of our Neural Slicer successes on all cages although the converging speed and the level of final loss are different. Caging meshes in different resolutions may have different genus numbers. 

The third limitation comes from how global collision is handled. When the collision between a printer-head and the model / platform occurs, we employ the same strategy as $S^3$-slicer~\cite{zhang2022s3} to increase the weight of the harmonic term to give more flat layers. In extreme cases, it will give planar layers that can guarantee no collision while other manufacturing objectives are compromised. The collision with robotic arms can be prevented in the downstream steps of motion planning~\cite{zhang2021singular,dai2020}.

\begin{figure}[t]
\centering
\includegraphics[width=\linewidth]{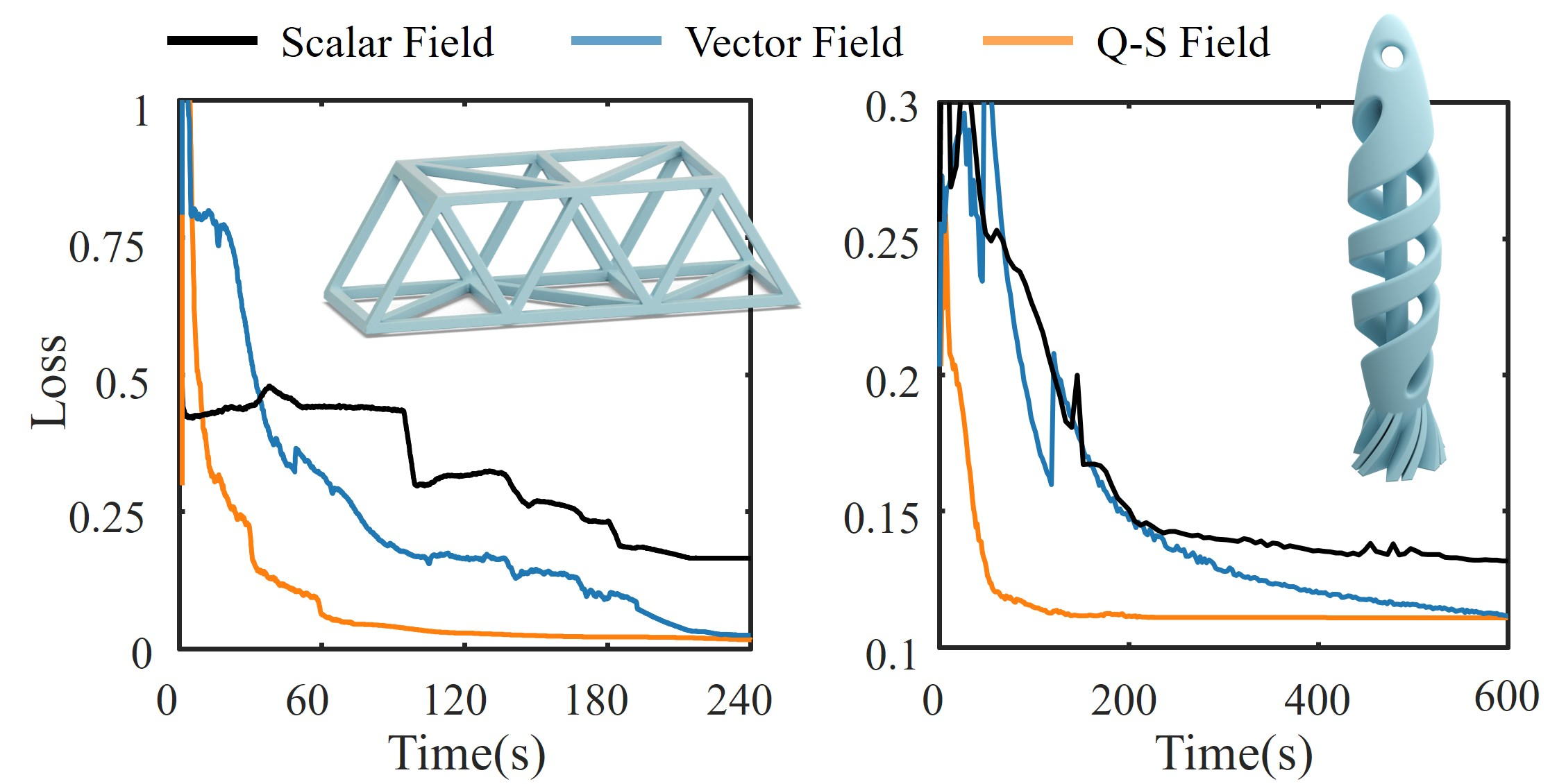}
    \put(-239,7){\small \color{black}(a)}
    \put(-116,7){\small \color{black}(b)}
\caption{Study of computational efficiency by using a scalar field, a vector field, and the quaternion-scaling field to parameterize the mapping, where the tests are conducted on (a) the Bridge model and (b) the Spiral Fish model. Clearly, the computation on quaternion-scaling field converges faster.}
\label{fig:QS_vs_vectorField}
\end{figure}

In our Neural Slicer, the mapping $\lambda(\cdot)$ is `parameterized' as a quaternion field $\mathbf{q}(\mathbf{x})$ and a scaling field $\mathbf{s}(\mathbf{x})$ (simply denoted as Q-S fields). In literature, shifting-vector based fields and scalar fields are widely used to represent a deformation field for non-rigid registration (ref.~\cite{deng2021deformed,sundararaman2022reduced,park2019deepsdf}. We argue that the deformation parameterized on the quaternion and the scaling fields is more efficient for computing a mapping for curved slicing. However, we are not able to prove this by theory. Comparisons are conducted on example models to study the computational efficiency by using different fields (see Fig.\ref{fig:QS_vs_vectorField}). 

\begin{figure}[t]
\centering
\includegraphics[width=\linewidth]{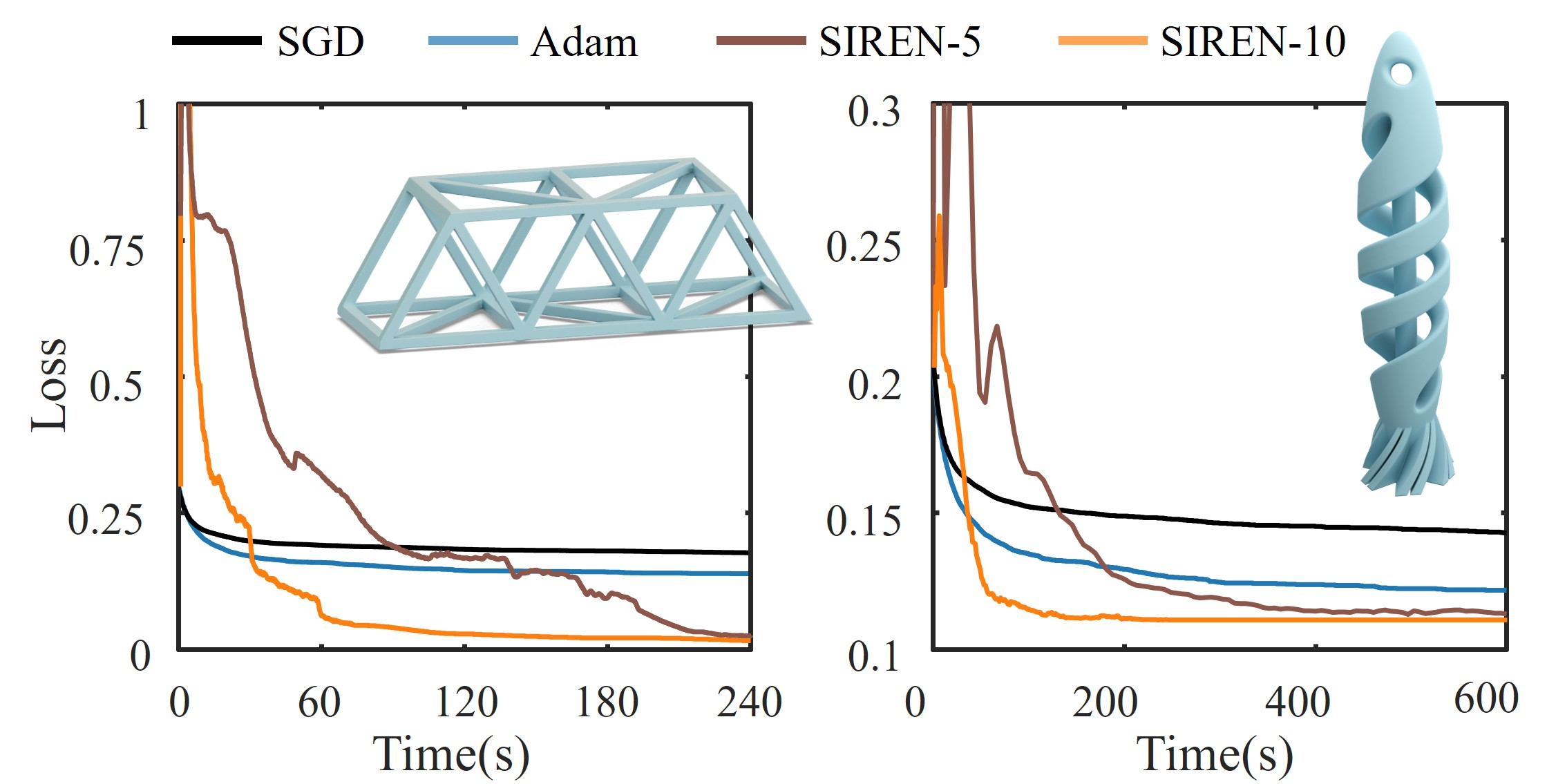}
    \put(-239,7){\small \color{black}(a)}
    \put(-116,7){\small \color{black}(b)}

\caption{Study of computational efficiency in directly using stochastic gradient descent (SGD) and adaptive moment estimation (Adam)~\cite{kingma2014adam} to optimize the quaternions and scales at the center of tetrahedral elements. The comparison with our approach using different SIREN layers (5 \& 10 layers) is also given on (a) the Bridge and (b) the Spiral Fish models.}
\label{fig:direct_opt}
\end{figure}

To demonstrate the effectiveness of our NN-based method to optimize the deformation as a mapping for slicing, we have conducted tests of directly using stochastic gradient descent (SGD) and Adaptive Moment Estimation (Adam)~\cite{kingma2014adam} to optimize the quaternions and scales at the center of tetrahedral elements and compared with our NN-based approach using different SIREN layers (5 \& 10 layers) on both the Bridge model and the Spiral Fish model (see Fig.\ref{fig:direct_opt} for the comparison). Significant differences in convergence speed can be observed -- especially when SIREN with 10 layers is employed. This is because such a neural network structure can capture and realize the required deformation more easily.

Explicitly imposing the requirement of injectivity in ARAP is not necessary in our framework. This is due to the fact that we directly evaluate manufacturing objectives as loss functions defined in the model space but not in the deformed space. The resultant $G(\mathbf{x})$ is a function mapping from $\mathbb{R}^3$ to $\mathbb{R}$ which allows the local rotation around the $z$-axis in the deformed space. This gives more flexibility to achieve optimal slicing results. The quality of resultant layers as isosurfaces are softly controlled by the harmonic terms. In examples with challenging geometry (e.g., the Spiral Fish model), we find that the allowed self-intersection is helpful to obtain results satisfying the manufacturing objectives better.

\section{Conclusion}
This paper introduces a representation-agnostic slicer for multi-axis 3D printing, utilizing a computational pipeline based on neural networks. Our slicer is designed to work on models with diverse representations and complex topology. We leverage a deformation field to compute the mapping, determining a scalar field in the space surrounding an input model. Isosurfaces of this scalar field are then extracted to generate curved layers for 3D printing. In contrast to previous approaches, where manufacturing objectives are indirectly optimized, our approach enables direct optimization of the scalar field by using loss functions directly based on local printing directions. This optimization is achieved through the implementation of a differentiable computation pipeline.
 Our approach demonstrates the successful handling of models with intricate topology, facilitated by the utilization of a volumetric mesh as a numerical computation cage. Notably, this allows for the input model and the cage to possess (i.e., different genus numbers). Leveraging the capabilities of a robust neural network solver, our approach allows for the flexible adjustment of local printing directions within a model, eliminating the need of good initial guesses for the optimization process. Our Neural Slicer presented in this paper can generate results with significantly enhanced performance, which has been verified in a variety of examples and in physical experiments.

\begin{acks}
The project is supported by the chair professorship fund at the University of Manchester and UK Engineering and Physical Sciences Research Council (EPSRC) Fellowship Grant (Ref.\#: EP/X032213/1).
\end{acks}

\bibliographystyle{ACM-Reference-Format}
\bibliography{reference.bib}

\end{document}